\documentclass[aps,prd,superscriptaddress,12pt,notitlepage]{revtex4}
    \usepackage{graphicx}
    \usepackage{amsmath,amssymb,mathrsfs}
\usepackage{epsf}
\usepackage{epsfig}
\usepackage[font=small,skip=0pt,justification=raggedright]{caption}

\newcounter{fig}   \newcommand{\lbfig}[1]{\refstepcounter{fig}
\label{#1} }

\newcommand{\Tr}{{\rm Tr}}

\newcommand{\bea}{\begin{eqnarray}}
\newcommand{\eea}{\end{eqnarray}}
\newcommand{\be}{\begin{equation}}
\newcommand{\ee}{\end{equation}}

\newcommand{\re}[1]{(\ref{#1})}

\newcommand{\tr}{\mbox{tr}}

\begin{document}

\title{Spinning gravitating Skyrmions in generalized Einstein-Skyrme model}

\author{I.~Perapechka}
\affiliation{Department of Theoretical Physics and Astrophysics, BSU, Minsk, Belarus}
\author{Ya.~Shnir}
\affiliation{Department of Theoretical Physics and Astrophysics, BSU, Minsk, Belarus\\
BLTP, JINR, Dubna, Moscow Region, Russia}

\begin{abstract}
We investigate properties of self-gravitating isorotating Skyrmions  in the generalized
Einstein-Skyrme model with higher-derivative terms in the matter field sector. These stationary
solutions are axially symmetric,
regular and asymptotically flat.  We provide a detailed account of the branch structure of the spinning
solutions in the topological sector of degree one.
We show that additional branches of solutions appear, as
the angular frequency increases above some critical value, these "cloudy" configurations can be considered as a bound system
of the spinning Skyrmions and "pion" excitations in the topologically trivial sector. Considering the critical behavior of the
isorotating solitons in the general Einstein-Skyrme model, we point out that there is no isospinning regular solutions
in the reduced self-dual $\mathcal{L}_6+\mathcal{L}_0$ submodel for any non zero value of the angular frequency.

\end{abstract}

\maketitle

\section{Introduction}
Various properties of self-gravitating solitons in 3+1 dimensional spacetime
have been under intense investigation over the last three decades.
Interesting examples are  static localized  field configurations
with finite energy in the Einstein-Skyrme theory \cite{Droz:1991cx,Bizon:1992gb},
Bartnik-McKinnon (BM) solutions in the $SU(2)$ Einstein-Yang-Mills model \cite{Bartnik:1988am}, and
gravitating monopoles in the $SU(2)$ Einstein-Yang-Mills-Higgs theory \cite{gmono}.
Further, it was discovered that many of those regular particle-like gravitating solitons, like for example
gravitating Skyrmions or monopoles, can be generalized to
contain inside the core a black hole of small horizon radius
\cite{Luckock:1986tr,Volkov:1989fi}.
These solutions can be viewed as bound states of Skyrmions and Schwarzschild black holes \cite{Ashtekar},
they provide counter-examples to the celebrated no-hair conjecture \cite{Ruffini:1971bza}.

Typically, there are two branches of self-gravitating regular static solitons. The lower in energy branch emerges
from the corresponding field configuration in the flat space, as the coupling to gravity is increasing. This branch of
solutions usually terminates at some critical value of the effective coupling, here it bifurcates with the second,
higher in energy branch which extends backwards as the coupling decreases. However, in the case of gravitating monopoles,
the second branch merges with the branch of extremal Reissner-Nordstr\"om black holes \cite{gmono} whereas
in the case of gravitating Skyrmions \cite{Droz:1991cx,Bizon:1992gb,Shnir:2015aba},
or axially-symmetric gravitating monopole-antimonopole configurations \cite{Kleihaus:2000hx,Kleihaus:2004fh}, the corresponding
upper branches of solutions extend back to the limit of vanishing coupling where they approach the lowest rescaled BM solution of the
$SU(2)$ Einstein-Yang-Mills theory. Typically, the upper branch solutions are unstable  \cite{Bizon:1992gb,Maeda:1993ap}.

Regular self-gravitating solitons of another type are stationary spinning field configurations, like boson stars
\cite{Friedberg:1986tp,Kleihaus:2005me} which in the flat space limit are linked to the non-topological Q-balls
\cite{Friedberg:1976me,Coleman:1985ki}, or rotating electrically charged sphalerons in the Einstein-Yang-Mills-Higgs theory
\cite{Kleihaus:2005fs}. While in the flat space  multimonopole configurations always possess zero angular momentum
\cite{Heusler:1998ec,VanderBij:2001nm}, there are spinning
excitations of classical Skyrmions  \cite{Battye:2005nx,Battye:2014qva}. On the other hand, in the context of application of the
Skyrme model to the nuclear physics, the semiclassical quantization
of angular momenta of spinning Skyrmion provides a natural way to identify the quantum numbers of a baryon \cite{Adkins:1983hy}.
Further, rotating Skyrmions  persist when gravity is coupled to the usual Skyrme model \cite{Ioannidou:2006nn}. It was observed
that the structure of the corresponding branches of solution becomes more complicated, there are additional branches of solutions,
which are not linked to the flat space Skyrmions or to the BM solutions  \cite{Ioannidou:2006nn}.

An interesting generalization of the Skyrme model in the flat space
was suggested recently to construct weakly bounded multisoliton configurations
\cite{Adam:2010fg,Adam:2010ds}. The Lagrangian of this generalized model contains an
additional sextic in derivatives term, it allows truncation of the model
to its limiting form, which supports self-dual equations. On the other hand the
sextic term in the general model is just the square of the topological current, it provides an additional repulsion
in the system which becomes important at high pressures or densities \cite{Adam:2015lra,Perapechka:2017yyc}.
Coupling generalized
Skyrme model to gravity provides a natural approximation to various
properties of neutron stars.

Investigation of the self-gravitating static solutions of the
general Einstein-Skyrme model reveals that the pattern of evolution along the branches becomes different from that
in the usual case. There always is
the stable lower branch of solutions, linked to the corresponding solitons in the flat space, however
the upper branch now terminates at a singular solution \cite{Adam:2016vzf}. Further, the presence of the Skyrme term is
a necessary condition for the existence of black holes with general Skyrmionic hair, both
in the asymptotically flat spacetime \cite{Adam:2016vzf,Gudnason:2016kuu} and in asymptotically AdS spacetime \cite{Perapechka:2016cof}.

In this paper we are considering properties of
stationary spinning solutions of the general Einstein-Skyrme model. We explore
axially symmetric configurations of the metric and matter fields and investigate their dependency both
on angular frequency and the effective gravitational coupling.

We confirm that similar to the case of spinning solitons in the usual Einstein-Skyrme model \cite{Ioannidou:2006nn},
additional branches of solutions appear in the generalized model with a sextic in derivatives term in
the matter field sector.
As the angular frequency increases above some critical value, the usual
branches of solutions merge new
 "cloudy" branches, these configurations can be considered as a bound system
of the spinning Skyrmions and "pion" excitations in the topologically trivial sector.
Considering the critical behavior of the
isorotating solitons in the general Einstein-Skyrme model, we found that there is no isospinning regular solutions
in the reduced self-dual $\mathcal{L}_6+\mathcal{L}_0$ submodel for any non zero value of the angular frequency.

The rest of the paper is structured as follows. In the next section we
briefly review the general Einstein-Skyrme model in the asymptotically flat spacetime.
Numerical results are presented in Section 3, where we
consider various patterns of the evolution of the stationary spinning configurations. For
the sake of compactness, we restrict the analysis to the simplest soliton  with
topological charge one. Conclusions and remarks are
formulated in Section 4.

\section{Generalized Einstein-Skyrme model}
The generalized Einstein--Skyrme model in asymptotically flat 3+1 dimensional space is defined by the
action
\be
\label{action}
S=\int{\sqrt{-g}\left(\frac{R}{16 \pi G}+\mathcal{L}\right)d^4 x},
\ee
where the gravity part of the action is the usual Einstein--Hilbert
action with curvature scalar $R$, $g$ denotes the determinant
of the metric, $G$ is the Newton gravitational  constant and the
matter part of the action is given by the Lagrangian $\mathcal{L}$
\be
\label{lag}
\mathcal{L}=\mathcal{L}_2+\mathcal{L}_4+\mathcal{L}_6+\mathcal{L}_0,
\ee
where $\mathcal{L}_0=\mu^2\mathcal{V}$ is a potential term with a mass parameter $\mu^2$,
\be
\label{l2l4}
\mathcal{L}_2=\frac{a}{2}\;\Tr\left(L_\mu L^\mu\right), \quad \mathcal{L}_4=\frac{b}{16}\;\Tr\left(\left[L_\mu,L_\nu\right]\left[L^\mu,L^\nu\right]\right)
\ee
are the usual kinetic term and the Skyrme term, respectively. Here $a$ and $b$ are nonnegative coupling constants and
\be
\label{cur}
L_\mu=U^\dagger \partial_\mu U
\ee
is the $\mathfrak{su}(2)$-valued left-invariant current, associated with the $\mbox{SU}(2)$-valued field
$$
U = \sigma \cdot {\mathbb I} + i \pi_a \cdot \tau^a ~~ _{\overrightarrow{r\to \infty}}~~~~
\mathbb{I} \, ,
$$
here $\tau^a$ are the usual
Pauli matrices. The quartet of the fields $(\sigma, \pi_a)$ is restricted to the surface
of the unit sphere, $\sigma^2+ \pi_a \cdot \pi_a =1$, thus, the field is a map from
compactified coordinate space $S^3$ to the $SU(2)$ group space, which is
isomorphic to the sphere $S^3$. The mapping is labeled by the topological invariant $B=\pi_3(S^3)$. Explicitly,
\be
\label{topcharge}
B=\frac{1}{24\pi^2}\int d^3x \varepsilon^{ijk}~\tr \left[(U^\dagger \partial_i U)
(U^\dagger \partial_j U)
(U^\dagger \partial_k U)\right] = \int d^3x \sqrt{-g} B_0\, ,
\ee
where integration is performed over three-dimensional hypersurface of constant temporal coordinate,
and $B_0$ is the temporal component of the topological current
\be
\label{topcur}
B^\mu=\frac{1}{24\pi^2\sqrt{-g}}\varepsilon^{\mu\nu\rho\sigma}\Tr\left(L_\nu L_\rho L_\sigma \right).
\ee
The Lagrangian density of the matter fields \re{lag} also includes the sextic term,
which is given by the square of the topological current:
\be
\label{l6}
\mathcal{L}_6=4\pi^4 c B_\mu B^\mu,
\ee
where the coupling $c$ is another nonnegative parameter of the general model.
Finally, the model includes the potential term $\mathcal{L}_0$, which is necessary to stabilize spinning solitons.
Here, for the sake of simplicity we consider the double-vacuum potential
\be
\label{doublevac}
\mathcal{V}=\Tr\left(\frac{\mathbb{I}+U}{2}\right)\Tr\left(\frac{\mathbb{I}-U}{2}\right) = 1-\sigma^2.
\ee

Introducing the dimensionless radial coordinate $\tilde r =\sqrt{\frac{a}{b}} r $ and the effective
gravitational coupling constant $\alpha^2={4\pi G a }$, we can rescale the parameters of the model \re{action} as
\be
\label{scaling}
a \to 1;~~ b \to 1;~~ c \to \tilde c =\frac{ac}{b^2}; ~~ \mu^2 \to \tilde \mu^2= \frac{b \mu^2}{a^2} \, .
\ee
Note that, similar to the usual Skyrme model, the limit $\alpha^2 \to 0$ can be approached in two different situations, namely
in the flat space limit, when the Newton constant $G \to 0$, or if the coupling constant $a \to 0$. Hence the pattern of evolution of the
regular self-gravitating solutions of the general Einstein-Skyrme model contains two branches \cite{Adam:2016vzf}.

Stationary spinning solutions of the model \re{action}
can be constructed by analogy with the usual Einstein-Skyrme model \cite{Ioannidou:2006nn}.
Using the axially symmetric time dependent parametrization of the Skyrme fields
\cite{Battye:2005nx,Battye:2014qva}
\be
\label{field}
\pi_1= \phi_1 \cos(n \varphi+\omega t);\quad \pi_2=\phi_1 \sin(n \varphi+\omega t);
\quad \pi_3=\phi_2;\quad \sigma = \phi_3
\ee
where $\phi_a$ is a triplet of field variables on the unit
sphere\footnote{Note that the parametrization \re{field} corresponds to isorotations of the pion field. However
our consideration is restricted to the configuration of topological degree one, thus
we can identify rotational and isorotational angular frequencies.},
we can now take into account the deformations of a spinning charge one
Skyrmion. Further, we  fix the winding $n=1$.

Note that the spinning configurations exist
for some set of values of the angular frequency up to $\omega_{max}$, for the particular choice of the potential
\re{doublevac} the configuration becomes unstable at $\omega_{max}=1$.

Indeed, in the flat space limit the stationary Lagrangian of the spinning configuration can be written as
\be
\int\mathcal{L}\;d^3 x = \frac12 \Lambda \omega^2 - M,
\ee
where $M$ is the classical mass of the Skyrmion, which is defined
as the spacial integral over the total energy density of the static configuration,
and $\Lambda$ is the moment of inertia about the $z$-axis:
\be
\Lambda = 4\pi \int\limits_0^\infty dr \int\limits_0^\pi d\theta r^2 \sin\theta(\lambda_2 + \lambda_4 + \lambda_6)
\label{inertia}
\ee
where the contributions from the terms $\mathcal{L}_2$, $\mathcal{L}_4$ and $\mathcal{L}_6$ in \re{lag} are
\be
\begin{split}
\lambda_2 &= -\phi_1^2; \qquad \lambda_4 = -\phi_1^2\left((\partial_r \phi_a)^2 + \frac{1}{r^2}
(\partial_\theta \phi_a)^2  \right);\\
\lambda_6 &=\frac{1}{r^2}\phi_1^2\biggl[
\phi_1(\partial_\theta\phi_2\partial_r\phi_3 - \partial_\theta\phi_3\partial_r\phi_2)\\
&+\phi_2(\partial_\theta\phi_3\partial_r\phi_1 - \partial_\theta\phi_1\partial_r\phi_3)
+\phi_3(\partial_\theta\phi_1\partial_r\phi_2 - \partial_\theta\phi_2\partial_r\phi_1)
\biggr]^2
\end{split}
\ee
respectively. In Fig.~\ref{Lambda} we presented the evaluated momenta of inertia of the Skyrmion, spinning about the third axis in the flat space,
as function of the angular frequency $\omega$ for a few values of the coupling $c$.  One can see that, as the angular frequency
approaches the critical value $\omega_{max}=1$, the momentum of inertia diverges for all values of the coupling $c$. The same observation holds
when the gravity is coupled to the generalized Skyrme field.

\begin{figure}[hbt]
\begin{center}
\includegraphics[height=.34\textheight, trim = 60 20 80 50, clip = true]{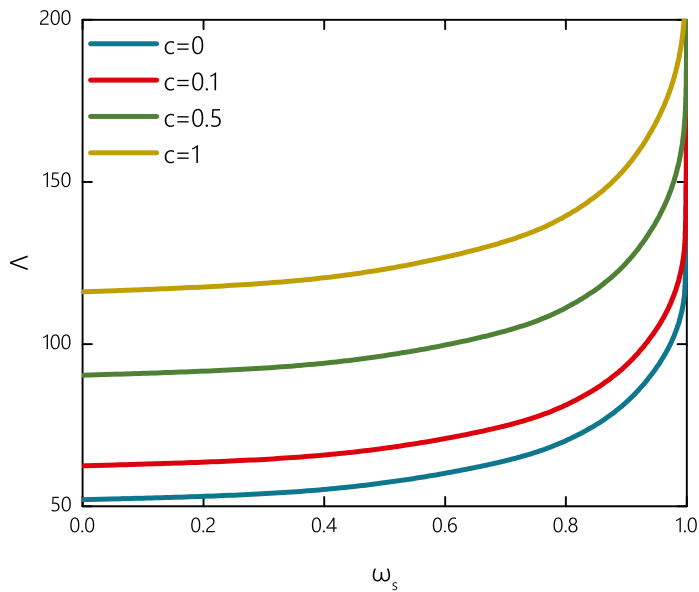}
\end{center}
\caption{\small
Momentum of inertia $\Lambda$ as a function of the angular frequency for $B = 1$
generalized Skyrmion isospinning about the $z$-axis
at $m = 1$ and a few values of coupling $c$.}
\lbfig{Lambda}
\end{figure}

Further, we employ the usual Lewis-Papapetrou metric in isotropic coordinates:
\be
\label{metric}
ds^2=-fdt^2+\frac{m}{f}\left(dr^2+r^2 d\theta^2\right)+\frac{l}{f}r^2\sin^2\theta\left(d\varphi-
\frac{o}{r}dt\right)^2,
\ee
where the metric functions
$f$, $m$, $l$ and $o$, as well as the matter fields $\phi^a$, are functions of
the radial variable $r$ and polar angle $\theta$, only. The $z$-axis ($\theta=0, \pi$) represents the symmetry axis.

As usually, total mass and the angular momentum of the stationary spinning axially symmetric
self-gravitating field configuration can be
evaluated from the boundary Komar integrals
\be
\label{mass}
M=-\frac{1}{2\alpha^2}\int\limits_S \nabla^\mu \xi^\nu dS_{\mu\nu} \, ,
\ee
and
\be
\label{angmom}
J=\frac{1}{4\alpha^2}\int\limits_S \nabla^\mu \eta^\nu dS_{\mu\nu},
\ee
where $\xi^\mu = (\partial_t,0,0,0), \eta^\mu=(0,0,0,\partial_\varphi)$
are two commuting Killing vector fields.
Alternatively,  these quantities can be expressed in terms of
the integrals over the three-dimensional space of the corresponding components of the
stress-energy tensor  $T_{\mu\nu}$, as
\be
\label{MJ}
\begin{split}
M=&2\int\limits_{\Sigma}\left(T_{\mu\nu}-\frac{1}{2}Tg_{\mu\nu}\right)k^\mu \xi^\nu dV =
\int(2 T_t^t- T)|g|^{1/2}~dr d\theta d\varphi;\\
J&=-\int\limits_{\Sigma}T_{\mu\nu}k^\mu \eta^\nu dV = -\int T_{\varphi}^t|g|^{1/2}~dr d\theta d\varphi \, ,
\end{split}
\ee
where $T=T^\mu_\mu$, $\Sigma$ is an asymptotically flat hyper-surface with a normal vector
$k_\mu = \frac{1}{\sqrt f}\left(1,0,0,\frac{o}{r}\right)$
and $dV= \frac{1}{\sqrt f} |g|^{1/2} dr d\theta d\varphi $ is the natural
volume element. We make use of these integrals to check our numerical results for consistency.

\section{Numerical results}
\subsection{Field equations and boundary conditions}
Variation of the rescaled action \re{action} with respect to the asymptotically flat metric $g_{\mu\nu}$ yields the Einstein equations
\be
R_{\mu\nu} -\frac12 R g_{\mu\nu} = 2\alpha^2 T_{\mu\nu}
\ee
where the matter field stress-energy tensor is
\be
\label{SET}
\begin{split}
T_{\mu\nu}&=\frac{2}{\sqrt{-g}}\frac{\partial\left(\mathcal{L}\sqrt{-g}\right)}{\partial g^{\mu\nu}}\\
&=-a\;\Tr\left(L_\mu L_\nu-\frac{1}{2}g_{\mu\nu}L_\rho L^\rho\right)-\frac{b}{4}\;\Tr\left(\left[L_\mu,L_\rho\right]
\left[L_\nu,L^\rho\right]-\frac{1}{4}g_{\mu\nu}\left[L_\rho,L_\sigma\right]\left[L^\rho,L^\sigma\right]\right)\\
&-8\pi^4 c\left(B_\mu B_\nu-\frac{1}{2}g_{\mu\nu}B_\rho B^\rho\right)+g_{\mu\nu}m_\pi^2\mathcal{V} \, .
\end{split}
\ee
Note that the complete system of stationary Einstein equations contains six nontrivial
equations for the four metric functions.  Following the usual approach suggested in
\cite{Kleihaus:1997mn,Ioannidou:2004mw} one can consider certain linear combinations of the Einstein equations, which
supplement the corresponding equations in the matter field sector. Alternatively one can substitute the
ansatz for the metric \re{metric} into the action \re{action}
and, eliminating the  total derivatives
of the metric functions, derive the corresponding system of variational equations.
One can check that the parametrization is consistent,
i.e. the  set of the equations, which follows from variation of the reduced stationary  action on
the ansatz \re{metric} coincides with the corresponding linear combinations of the Einstein equations.

The complete set of the axially symmetric field equations, which describes spinning field configurations in the
general Einstein-Skyrme model in addition also includes three equations on the matter
fields $\phi_a$.  Hence altogether we have a set of seven coupled elliptic
partial differential equations with mixed derivatives, to be solved numerically
subject to the appropriate boundary conditions.
As usual, they follow from the condition of regularity of the fields on the symmetry axis
and symmetry requirements as well as
the condition of finiteness of the energy of the system. In particular we have to take into account that
the asymptotic value of the Skyrme field is restricted to the vacuum and
the metric functions must approach unity at the spacial boundary.
Explicitly, we impose
\be
\label{BCorigin}
\begin{split}
\phi_1\bigl.\bigr|_{r=0}&=0,\quad \phi_2\bigl.\bigr|_{r=0}=0,\quad \phi_3\bigl.\bigr|_{r=0}=-1,\quad \\
\partial_r f\bigl.\bigr|_{r=0}&=0,\quad \partial_r l\bigl.\bigr|_{r=0}=0,\quad
\partial_r m\bigl.\bigr|_{r=0}=0,\quad o\bigl.\bigr|_{r=0}=0 \, ,
\end{split}
\ee
while in the sector of topological degree one  the boundary conditions on spatial infinity are:
\be
\label{BCinf}
\begin{split}
\phi_1\bigl.\bigr|_{r\rightarrow\infty}&\rightarrow 0,\quad \phi_2\bigl.\bigr|_{r\rightarrow\infty}\rightarrow 0,\quad
\phi_3\bigl.\bigr|_{r\rightarrow\infty}\rightarrow 1,\quad \\
f\bigl.\bigr|_{r\rightarrow\infty}&\rightarrow 1,\quad l\bigl.\bigr|_{r\rightarrow\infty}\rightarrow 1,\quad
m\bigl.\bigr|_{r\rightarrow\infty}\rightarrow 1,\quad o\bigl.\bigr|_{r\rightarrow\infty}\rightarrow 0 \, .
\end{split}
\ee
Boundary conditions on the symmetry axis are
\be
\label{BCz}
\begin{split}
\phi_1\bigl.\bigr|_{\theta=0}&=0,\quad \partial_\theta \phi_2\bigl.\bigr|_{\theta=0}=0,
\quad \partial_\theta \phi_3\bigl.\bigr|_{\theta=0}=0,\quad \\
\partial_{\theta} f\bigl.\bigr|_{\theta=0}&=0,\quad \partial_{\theta} l\bigl.\bigr|_{\theta=0}=0,\quad
\partial_{\theta} m\bigl.\bigr|_{\theta=0}=0,\quad \partial_{\theta} o\bigl.\bigr|_{\theta=0}=0,
\end{split}
\ee
Finally, the boundary conditions on the $xy$-plane follow from the reflection symmetry:
\be
\label{BCxy}
\begin{split}
\partial_\theta \phi_1\bigl.\bigr|_{\theta=\frac{\pi}{2}}&=0,\quad \phi_2\bigl.\bigr|_{\theta=\frac{\pi}{2}}=0,
\quad \partial_\theta
\phi_3\bigl.\bigr|_{\theta=\frac{\pi}{2}}=0,\quad \\
\partial_{\theta} f\bigl.\bigr|_{\theta=\frac{\pi}{2}}&=0,\quad \partial_{\theta} l\bigl.\bigr|_{\theta=\frac{\pi}{2}}=0,
\quad \partial_{\theta} m\bigl.\bigr|_{\theta=\frac{\pi}{2}}=0,\quad \partial_{\theta} o\bigl.\bigr|_{\theta=\frac{\pi}{2}}=0.
\end{split}
\ee
As usually, regularity on the $z$-axis requires $m\bigl.\bigr|_{\theta=0}=l\bigl.\bigr|_{\theta=0}$ \cite{Hartmann:2001ic}, so it is convenient
to introduce an auxiliary function $h = \frac{l}{m}$ imposing the following boundary condition on this function:
\be
\label{BCh}
h\bigl.\bigr|_{r=0}=1, \quad h\bigl.\bigr|_{r\rightarrow\infty}\rightarrow 1, \quad h\bigl.\bigr|_{\theta=0}=1, \quad \partial_{\theta} h\bigl.\bigr|_{\theta=\frac{\pi}{2}}=0.
\ee

\begin{figure}[hbt]
\begin{center}
\includegraphics[width=.46\textwidth, trim = 40 20 90 20, clip = true]{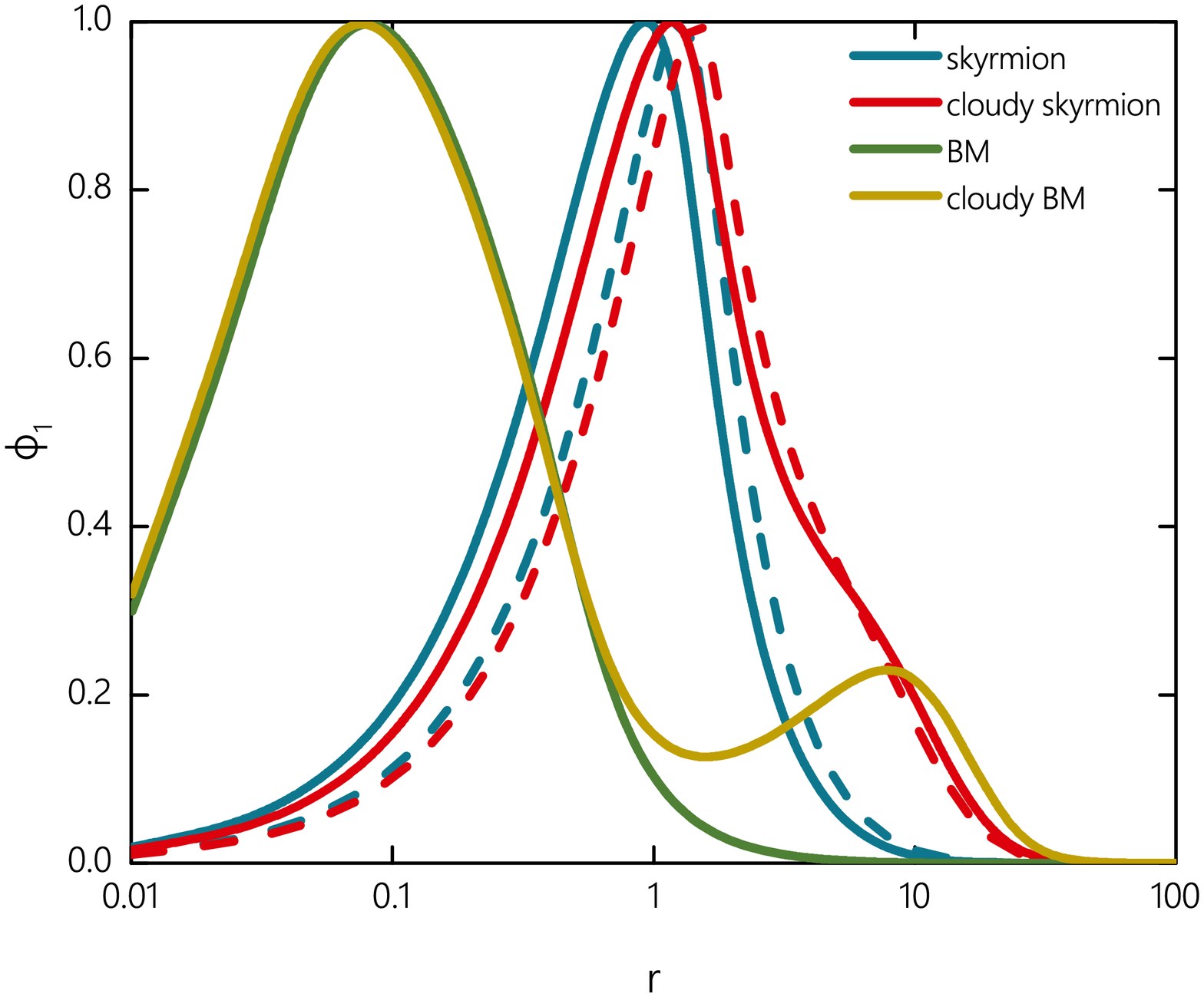}
\includegraphics[width=.46\textwidth, trim = 40 20 90 20, clip = true]{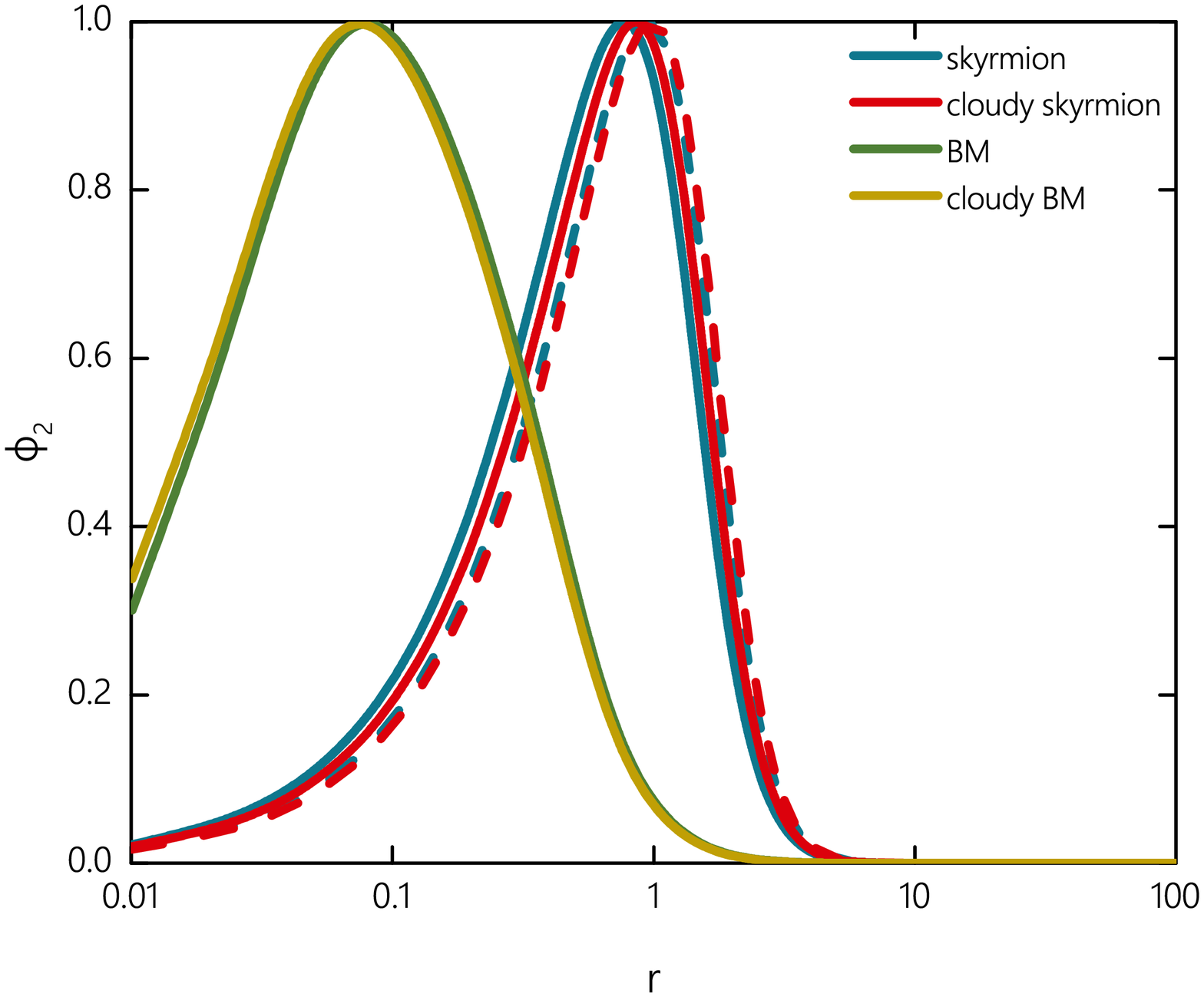}
\includegraphics[width=.46\textwidth, trim = 40 20 90 20, clip = true]{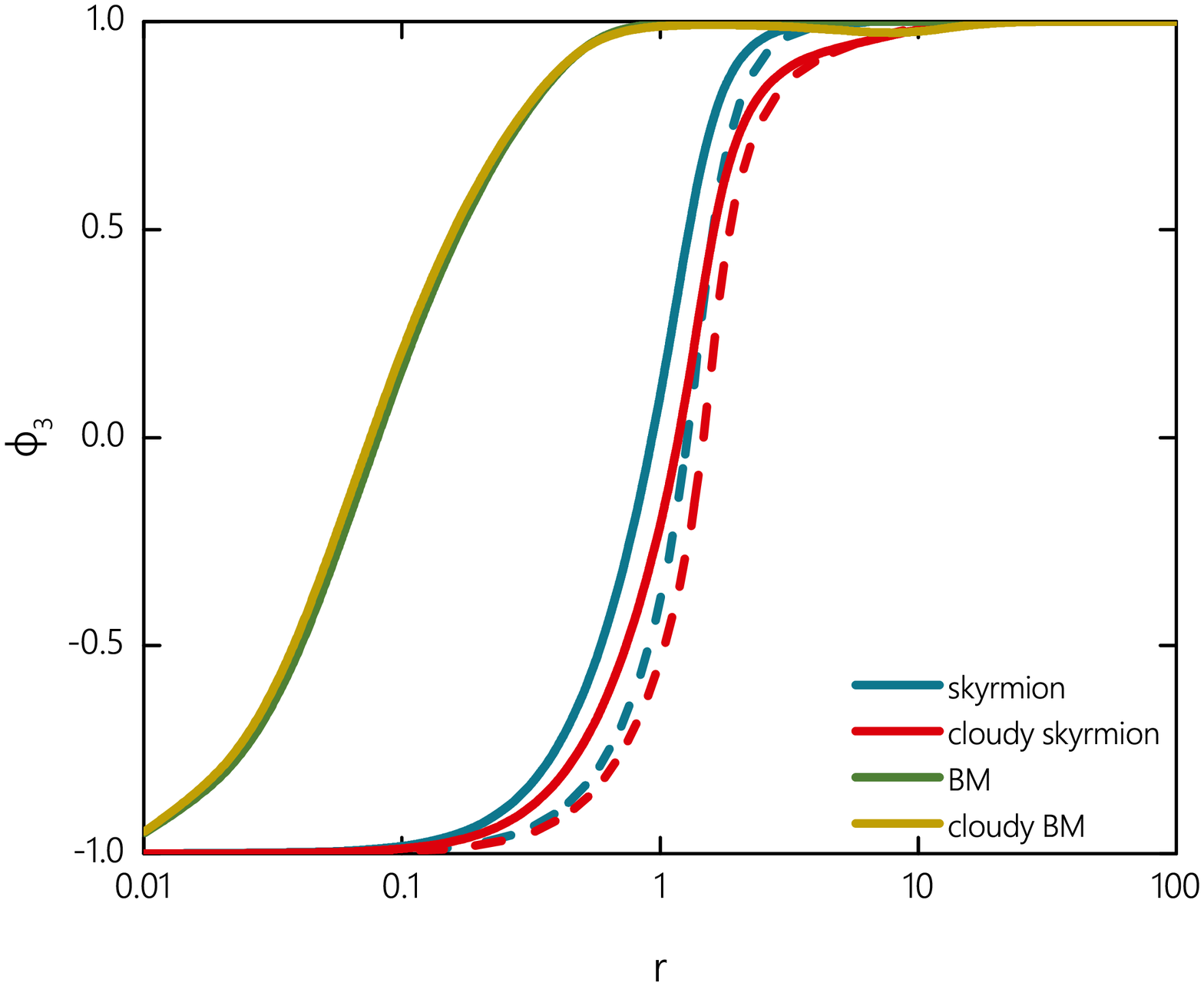}
\end{center}
\caption{\small
The profiles of the Skyrmion functions $\phi_1$ at $\theta=\pi/2$ (upper left), $\phi_2$ at $\theta=0$(upper right)
and $\phi_3$ at $\theta=\pi/2$ (bottom) for $c=0$
(solid lines) and $c=1$ (dashed lines) are plotted for different types of configurations
at $\omega=0.97$ and $\alpha=0.15$.}
\lbfig{3dfields}
\end{figure}

With these boundary conditions at hand we now can perform the integration over radial coordinate $r$ and
obtain the following expression for the total mass \re{mass} of the regular stationary spinning axially symmetric gravitating Skyrmion
\be
\label{massfinal}
M=\frac{1}{2G}\lim\limits_{r\rightarrow\infty}r^2 \partial_r f \, .
\ee
Similarly, for the total angular  momentum  of the configuration \re{angmom}, we find
\be
\label{angmomfinal}
J=\frac{1}{2G}\lim\limits_{r\rightarrow\infty}r^2  o \, .
\ee
Thus, the mass and the angular momentum of the configuration can be
read off the asymptotic expansion of the metric functions $f$ and $o$, respectively \cite{Kleihaus:2005me,Kleihaus:2000kg}
\be
f = 1-\frac{2MG}{r} +O\left(\frac{1}{r^2}\right)\, , \qquad
o= -\frac{2JG}{r^2} + O\left(\frac{1}{r^3}\right)\, .
\ee

\subsection{Numerical results and discussion}

To find solutions of the set of equations which follow from the action \re{action} and
depend parametrically on the effective gravitational constant $\alpha$ and on the angular frequency $\omega$,
we used the software package CADSOL based on the Newton-Raphson algorithm \cite{schoen}.
The numerical calculations are mainly performed on an equidistant grid
in spherical coordinates $r$ and $\theta$. Typical grids we used have sizes $55 \times 30$.
Here we map the infinite interval of the variable $r$ onto the compact radial coordinate
$x=\frac{r/r_0}{1+r/r_0}  \in [0:1]$.
The parameter $r_0$ is used to improve the accuracy of numerical solution.
Constraint of the Skyrme fields to the surface of unit sphere $\phi_a \phi_a=1$,
is implemented via inclusion in the Lagrangian term $k(\phi_a\phi_a-1)^2$, where $k$ is
a suitable Lagrange multiplier. Anywhere, where converse is not stated explicitly, we fix $a=b=\mu=1$.

\begin{figure}[hbt]
\lbfig{3dX}
\begin{center}
\includegraphics[width=.49\textwidth, trim = 30 30 100 20, clip = true]{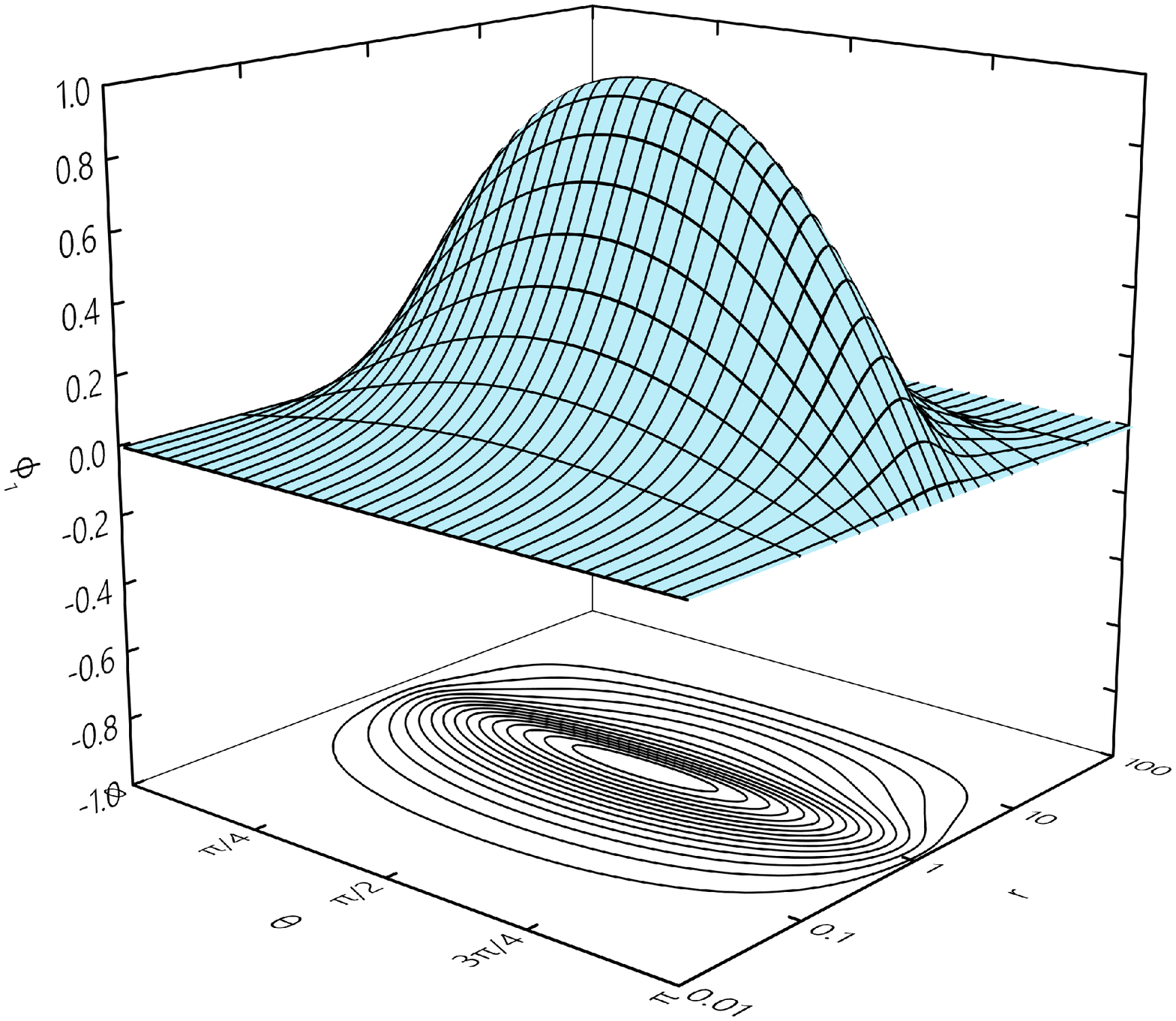}
\includegraphics[width=.49\textwidth, trim = 30 30 100 20, clip = true]{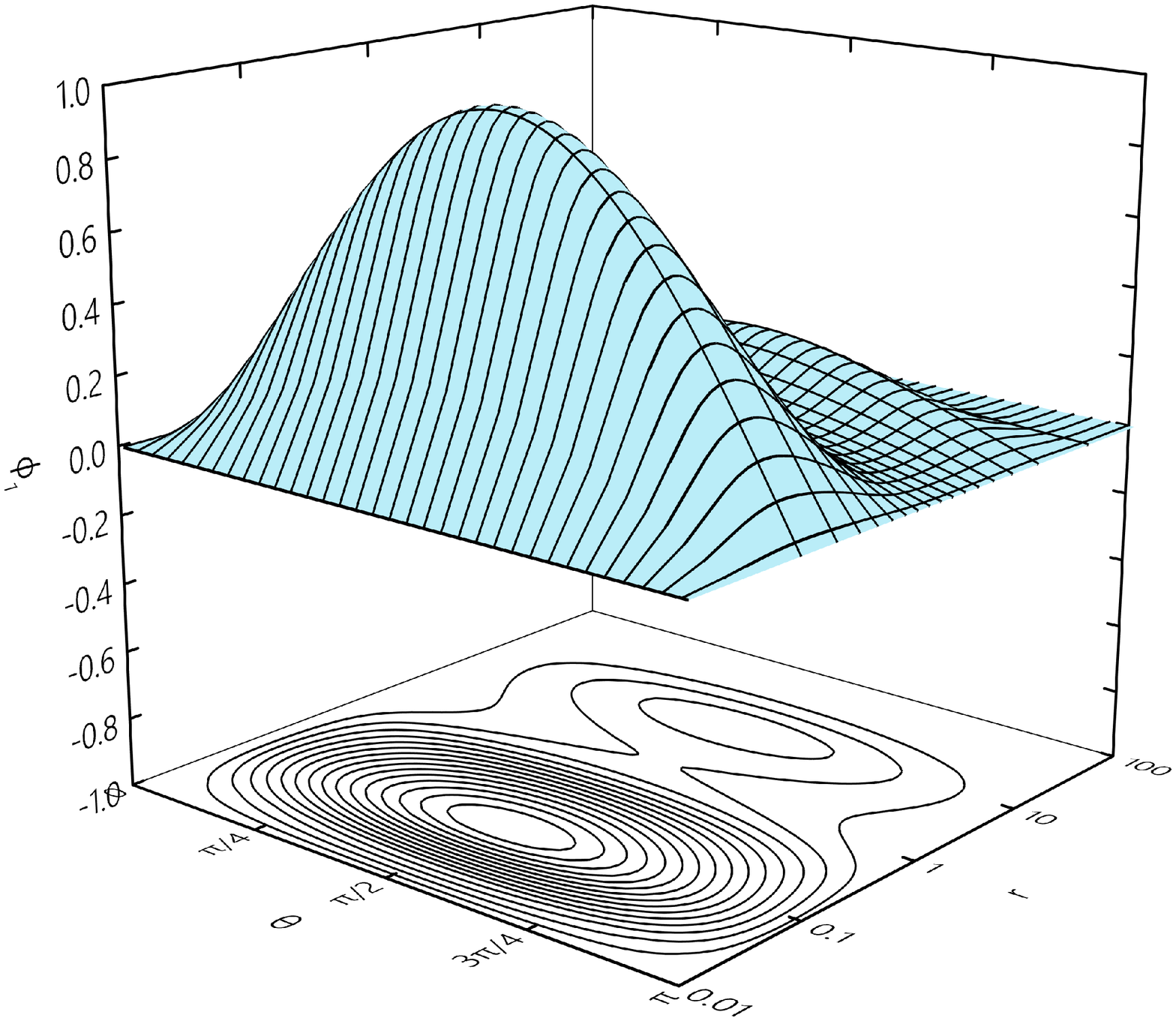}
\end{center}
\caption{\small
Distributions of Skyrme field component $\phi_1$ for `cloudy skyrmion' (left panel) and `cloudy BM' (right panel) solution with $c=0,\alpha=0.15, \omega=0.97$.
}
\end{figure}

Spinning gravitating solution of the usual Einstein-Skyrme model have been studied before \cite{Ioannidou:2006nn}.
The general pattern is that there always are two branches of solutions which,
in the limit of vanishing effective gravitational coupling constant $\alpha$,
tend to the flat space Skyrmions and to the rescaled BM solutions, respectively. It was suggested to refer to these
branches to as the `skyrmion' branch and `BM' branch, respectively  \cite{Ioannidou:2006nn}. As the angular frequency
$\omega$ remains much smaller than $\omega_{max}=1$, these two branches merge at some critical value of the effective
gravitational coupling $\alpha_{cr}$, as it happens in the case of the usual self-gravitating Skyrmions
\cite{Droz:1991cx,Bizon:1992gb}. The critical value $\alpha_{cr}$ increases with  increasing $\omega$.
Unlike the static self-gravitating solutions, stationary spinning Skyrmions exhibit a loop-like
gravitational coupling dependence of the mass, see Fig.~\ref{constomega}, upper left plot.

\begin{figure}[hbt]
\lbfig{3dY}
\begin{center}
\includegraphics[width=.49\textwidth, trim = 30 30 100 20, clip = true]{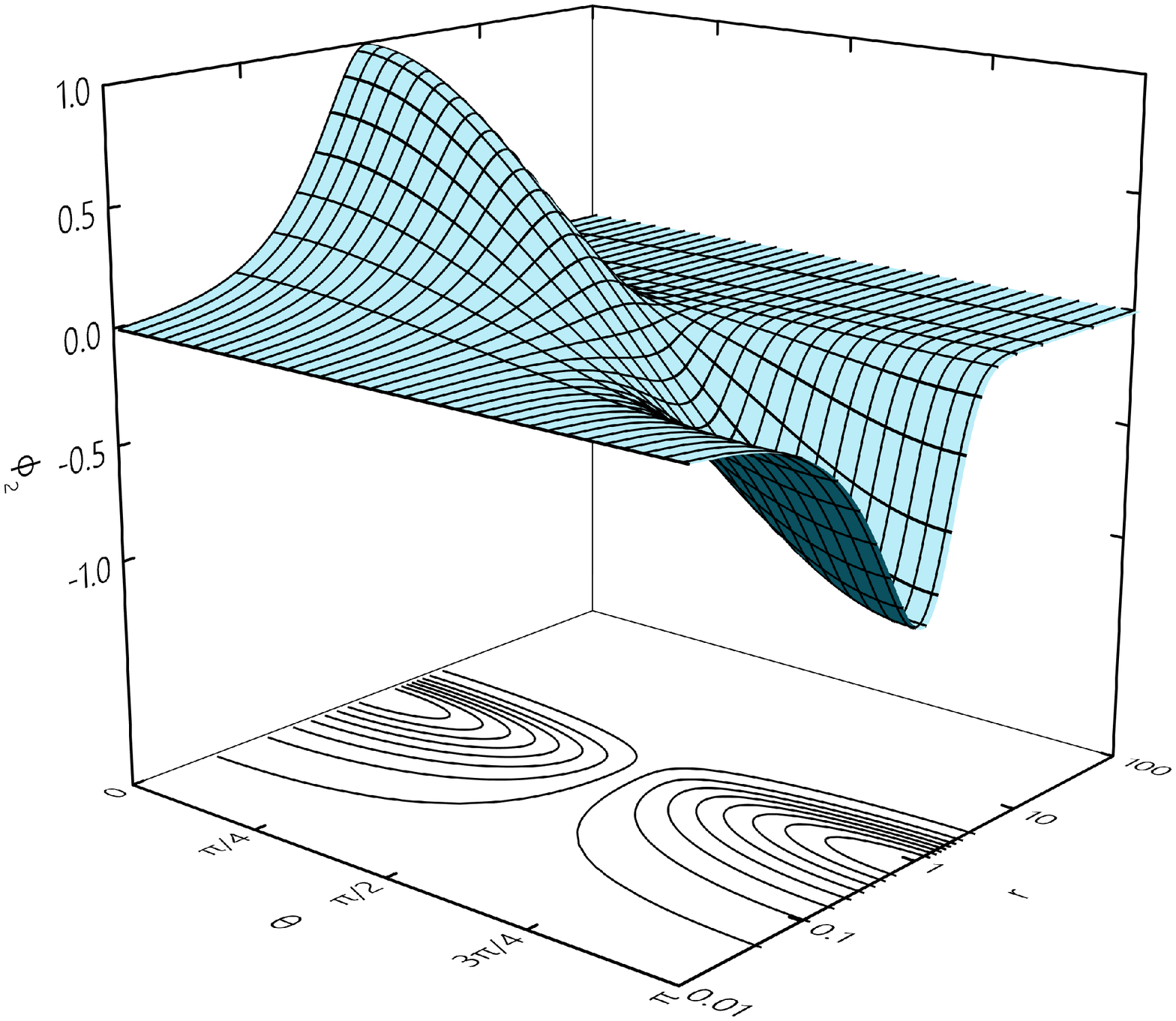}
\includegraphics[width=.49\textwidth, trim = 30 30 100 20, clip = true]{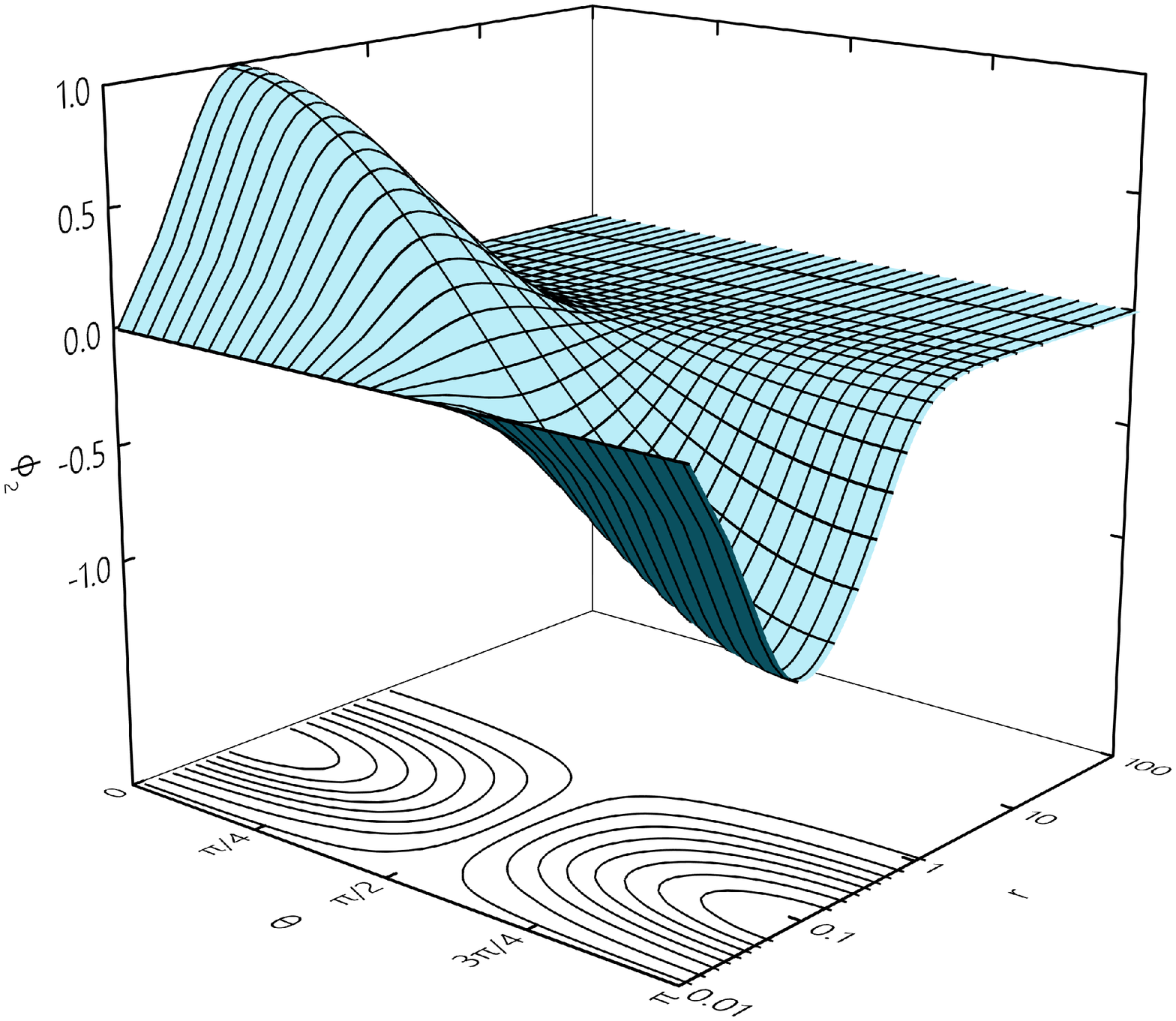}
\end{center}
\caption{\small
Distributions of Skyrme field component $\phi_2$ for `cloudy skyrmion' (left panel) and `cloudy BM' (right panel) solution with $c=0,\alpha=0.15, \omega=0.97$.
}
\end{figure}

The pattern becomes more complicated as $\omega$ approaches the second critical value $\omega_{cr} < \omega_{max} =1$.
If the gravitational coupling is relatively large, new
branches of solutions of different type appear, they form so called `cloudy branches' \cite{Ioannidou:2006nn}. These
field configurations, bounded by strong gravitational attraction, resembles, to a certain degree, the boson stars
\cite{Friedberg:1986tp,Kleihaus:2005me}. The clouds
possess zero topological charge, they do not exists in the limit $\alpha \to 0$, or if the angular frequency is small.
Pure pion clouds can be constructed via setting
\be
\phi_1 = \cos H(r,\theta),\qquad
\phi_2 = 0,\qquad
\phi_3 = \cos H(r,\theta),
\ee
where the profile function $H(r,\theta)$ vanishes on the symmetry axis and at spatial infinity. On the other hand, as the
angular frequency of the spinning gravitating Skyrmion approaches the critical value $\omega_{max}$,
the configuration behave like a compacton,
it becomes strongly localized within an interior region. In the outer region the fields are taking the vacuum values,
there the pion clouds
may appear as excitations \cite{Ioannidou:2006nn}. These field configurations can be considered as superposition of the
spinning Skyrmions both on the upper and on the lower branches, and pion clouds.
This pattern of course,
depends on the explicit form of the potential term and on the value of the mass parameter $m$.
In Figs.~\ref{3dX}--\ref{3do} we present different types of the solutions.

\begin{figure}[hbt]
\lbfig{3dZ}
\begin{center}
\includegraphics[width=.49\textwidth, trim = 30 30 100 20, clip = true]{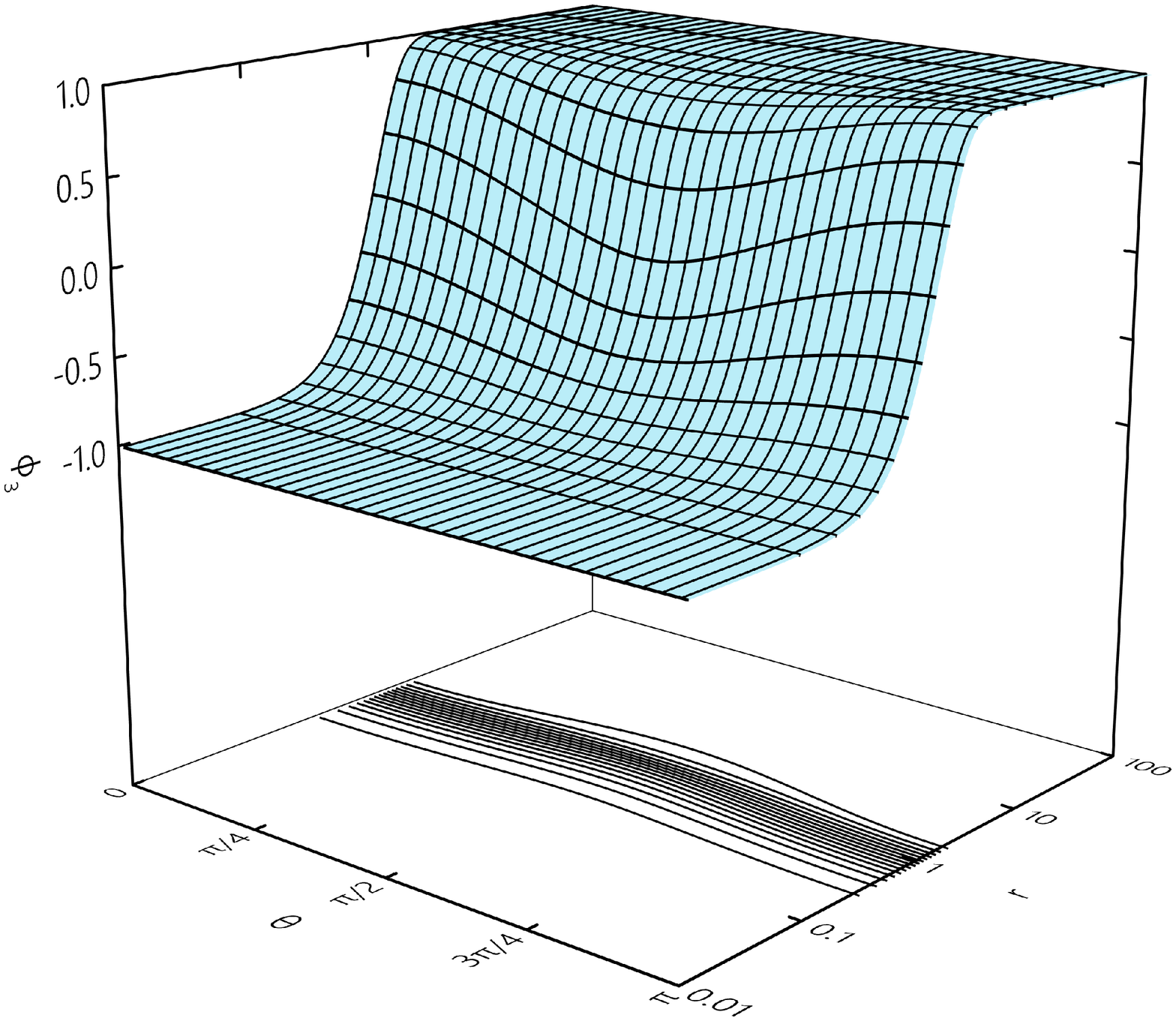}
\includegraphics[width=.49\textwidth, trim = 30 30 100 20, clip = true]{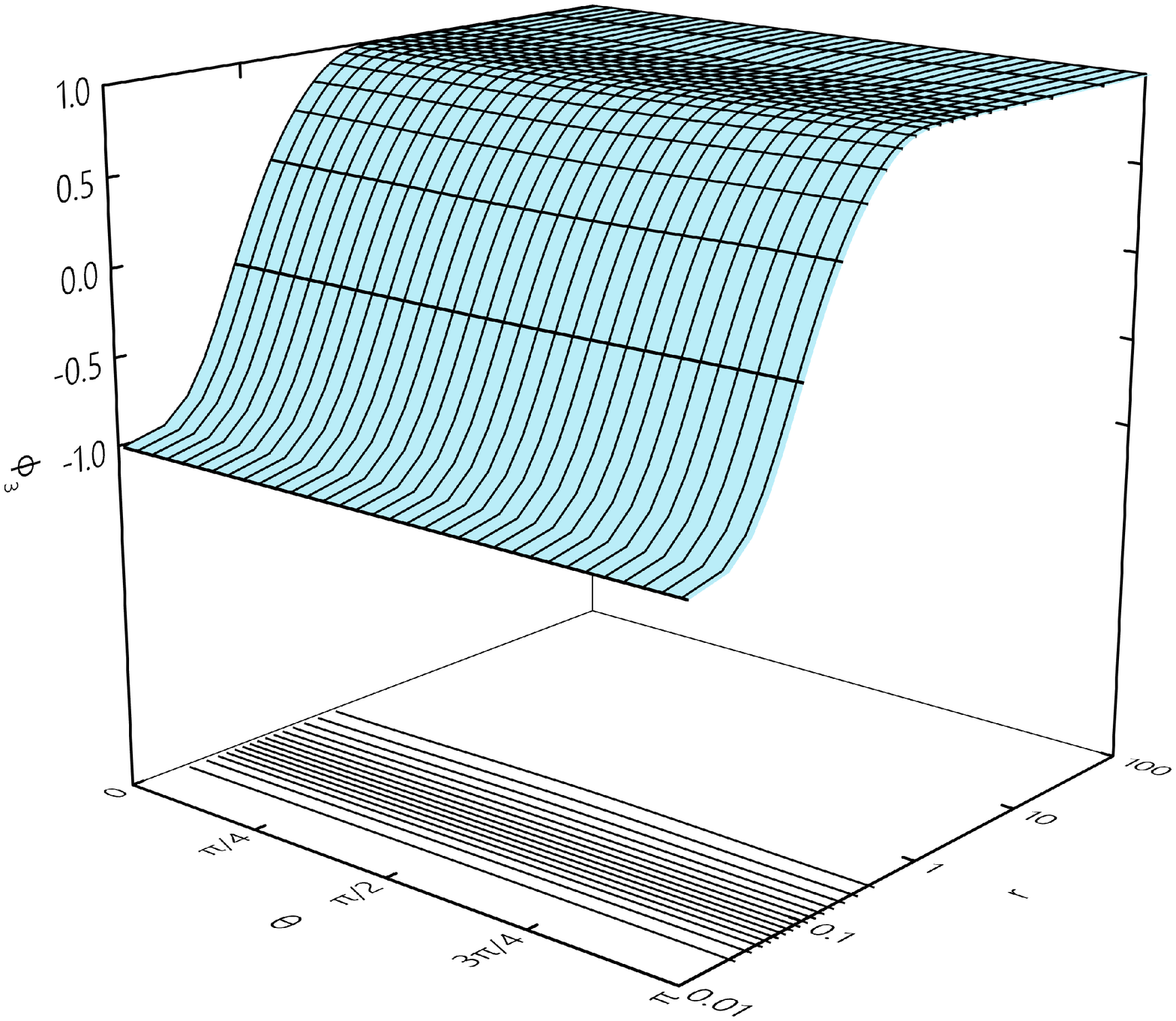}
\end{center}
\caption{\small
Distributions of Skyrme field component $\phi_3$ for `cloudy skyrmion' (left panel) and `cloudy BM' (right panel) solution with $c=0,\alpha=0.15, \omega=0.97$.
}
\end{figure}

We can expect the similar set of solutions persists in the generalized Skyrme model \re{lag}. Firstly, we note that the
sextic term \re{l6} in the matter field Lagrangian is defined as the square of the topological current,
thus it does not affect the pion clouds since they are excitations in the topologically trivial sector.
Considering the branch structure of the rotating gravitating Skyrmions in the
generalized Skyrme model we observe that, for small-to-moderate values of
the angular frequency $\omega < \omega_{cr}$, the usual pattern remains, there is a branch of gravitating spinning
solitons, which originates from the corresponding flat space configurations
and another branch of regular solutions, which bifurcates with the first branch
at some critical value of the gravitational coupling $\alpha_{cr}$. The mass of the spinning Skyrmions decreases along
the first branch, while the angular momentum initially increases and then, as the gravitational coupling starts to approach
the critical value, begin to decrease, see the right upper plot in Fig.~\ref{constomega}.

\begin{figure}[hbt]
\lbfig{3df}
\begin{center}
\includegraphics[width=.49\textwidth, trim = 40 30 100 20, clip = true]{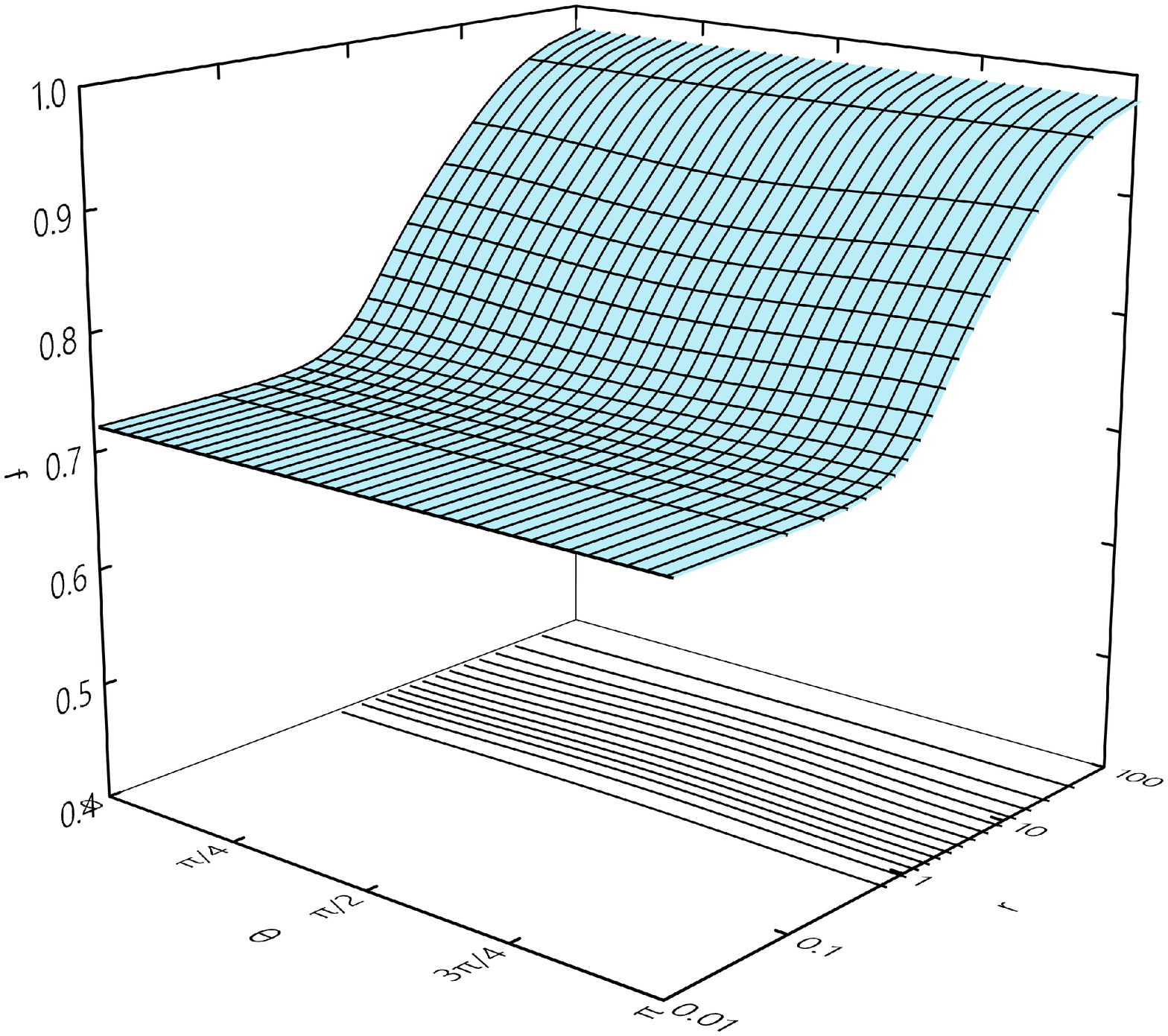}
\includegraphics[width=.49\textwidth, trim = 40 30 100 20, clip = true]{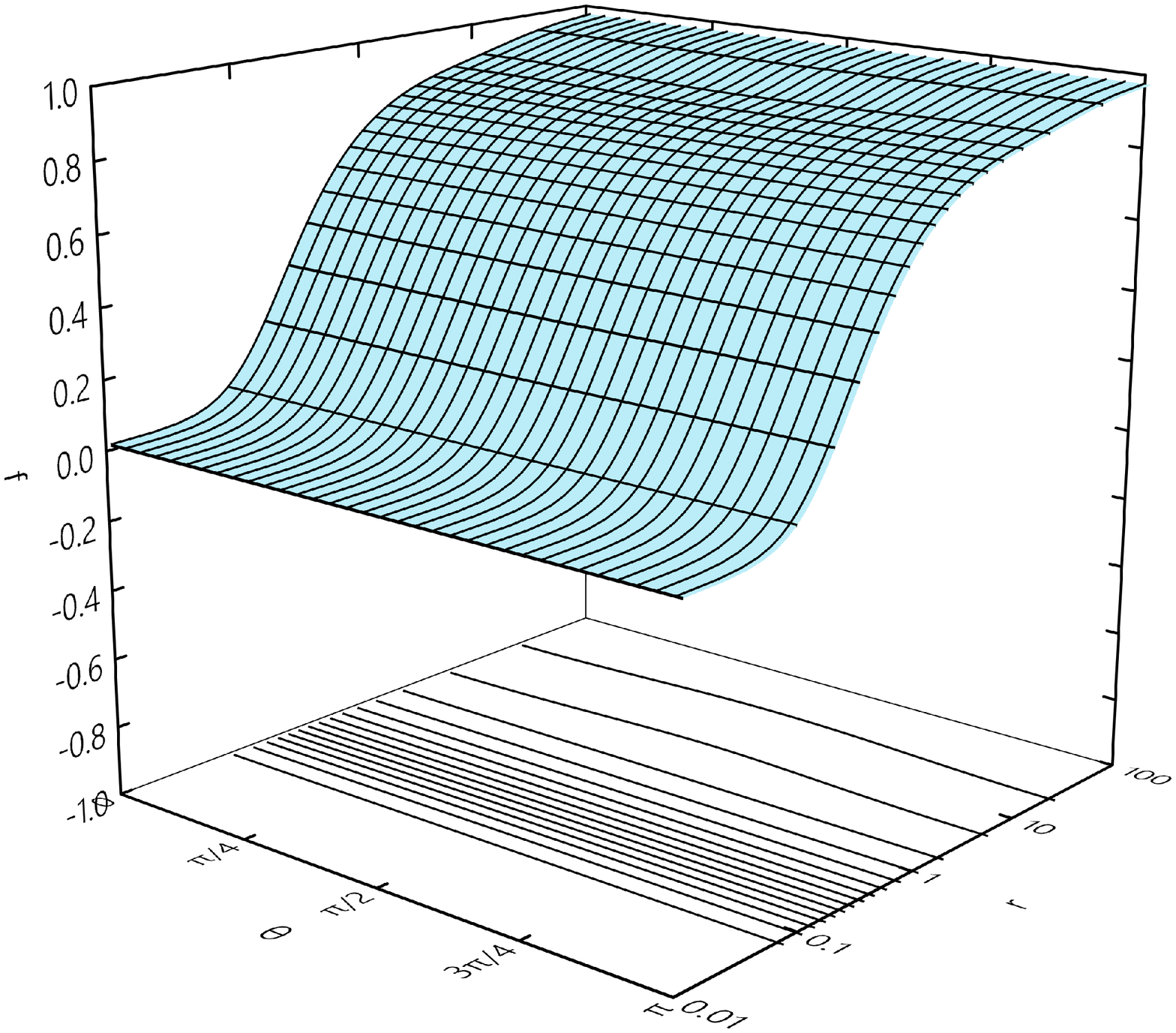}
\end{center}
\caption{\small
Distributions of metric function $f$ for `cloudy skyrmion' (left panel) and `cloudy BM' (right panel) solution with $c=0,\alpha=0.15, \omega=0.97$.
}
\end{figure}

Since the effective gravitational coupling is defined as $\alpha^2={4\pi G a }$, the evolution along
the second branch is related with decrease of the coupling constant $a$.
However, for any $c \neq 0$,
this branch is no longer linked to the Bartnik-McKinnon solution, it
exists only for $\alpha>\alpha_{min}$, at which $f(0) \to 0$,
both in the case of spinning and non-spinning configurations.
The value of $\alpha_{min}$ increases with increasing of $\omega$, further
as the coupling  $c$ to the sextic term  increases, both $\alpha_{min}$ and the value of
the critical coupling $\alpha_{cr}$ are also increasing, see
Fig.~\ref{constomega}.

\begin{figure}[hbt]
\lbfig{3dm}
\begin{center}
\includegraphics[width=.49\textwidth, trim = 40 30 100 20, clip = true]{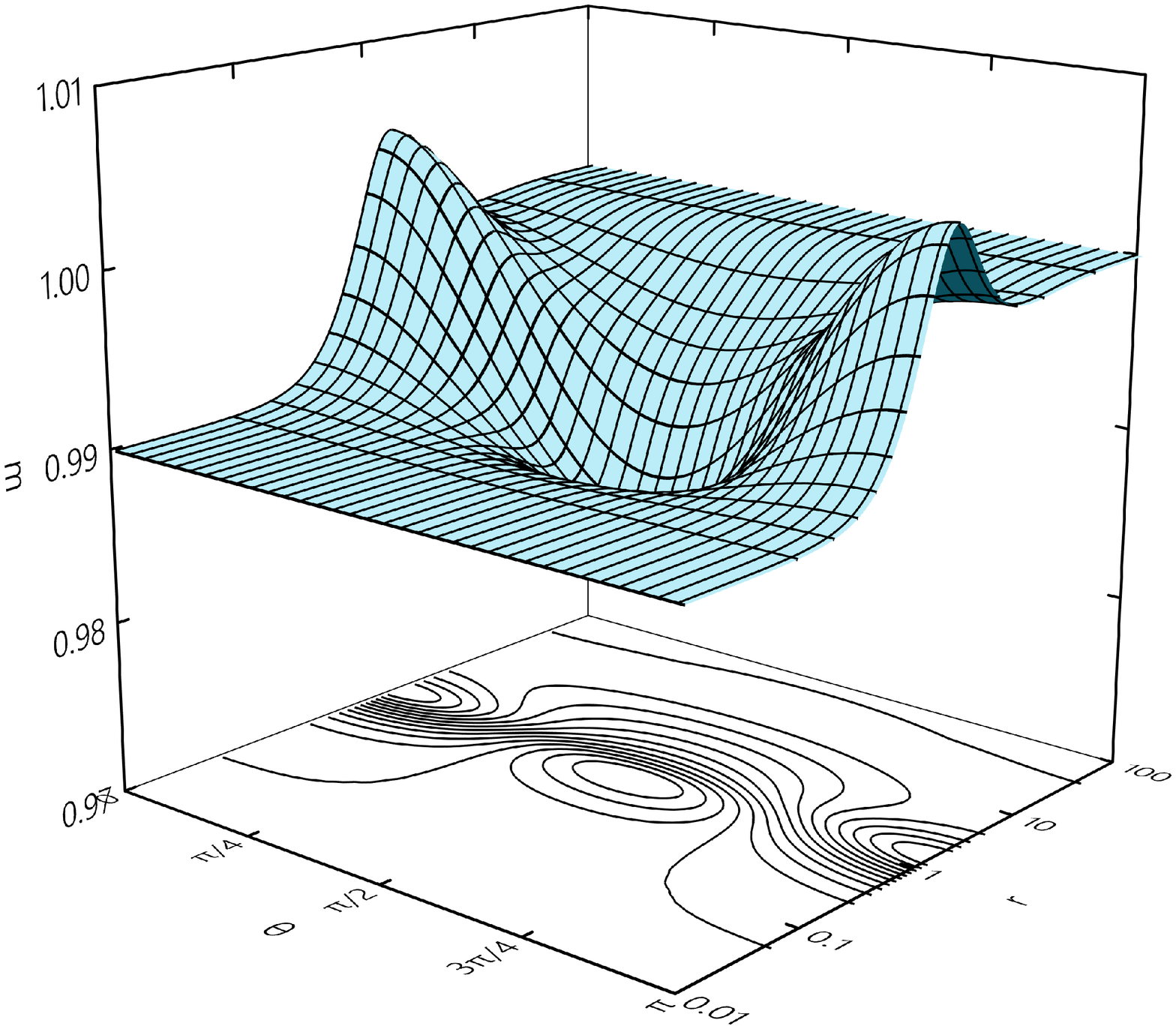}
\includegraphics[width=.49\textwidth, trim = 40 30 100 20, clip = true]{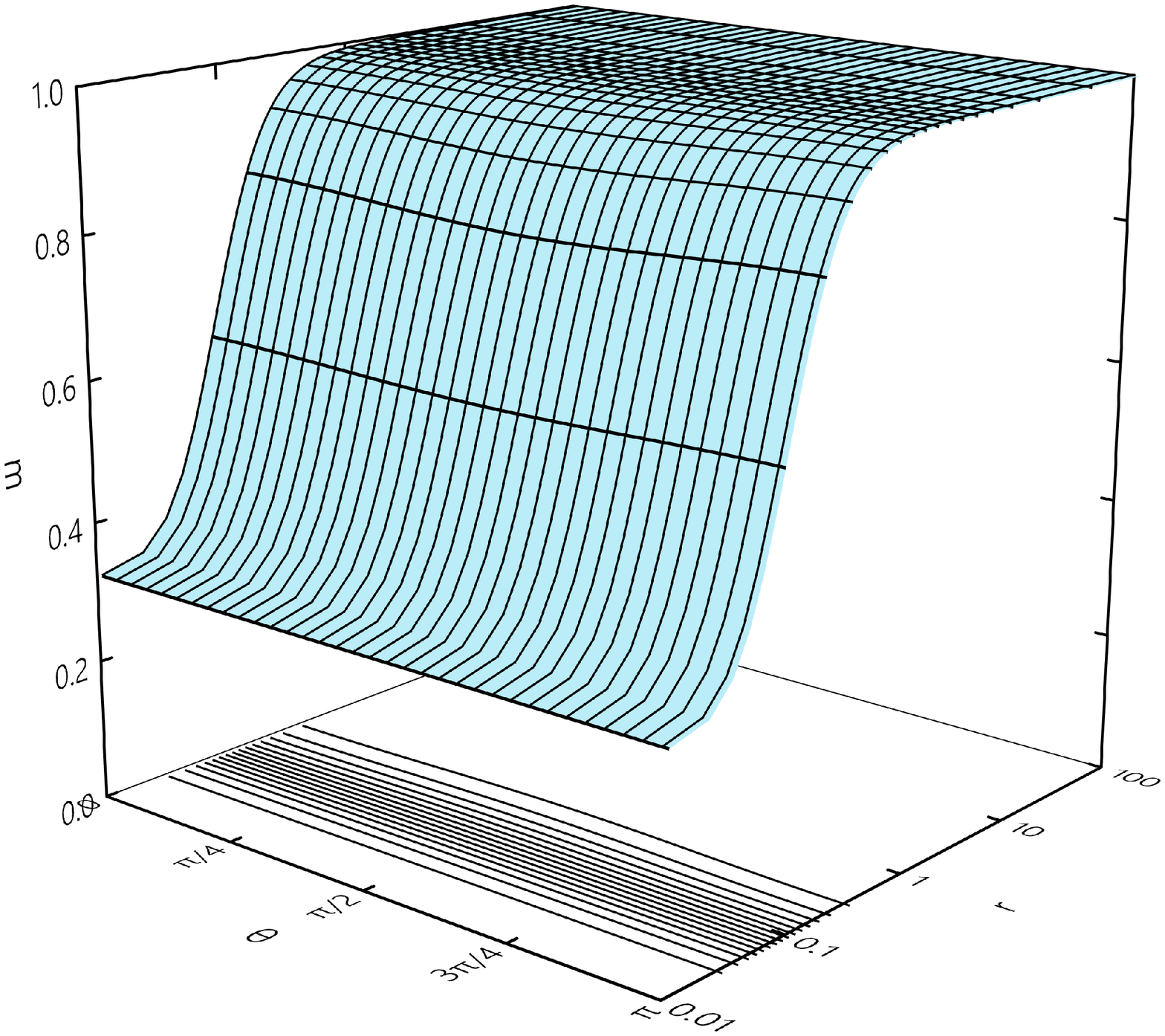}
\end{center}
\caption{\small
Distributions of metric function $m$ for `cloudy skyrmion' (left panel) and `cloudy BM' (right panel) solution with $c=0,\alpha=0.15, \omega=0.97$.
}
\end{figure}

The values of the metric functions at the origin, $f(0)$ and $l(0)$, both
decrease monotonically, first along the
`skyrmion' branch, and then, along the `BM' branch, as $\alpha$ decreases. In the general model
this branch terminates at a singular solution.
The angular momentum of spinning Skyrmions continues to decrease on the second branch, for $c \neq 0$ it
approaches its finite minimal value as the configuration tends to the singular solution, as seen in the right
upper plot, Fig.~\ref{constomega}

\begin{figure}[hbt]
\lbfig{3dl}
\begin{center}
\includegraphics[width=.49\textwidth, trim = 40 30 100 20, clip = true]{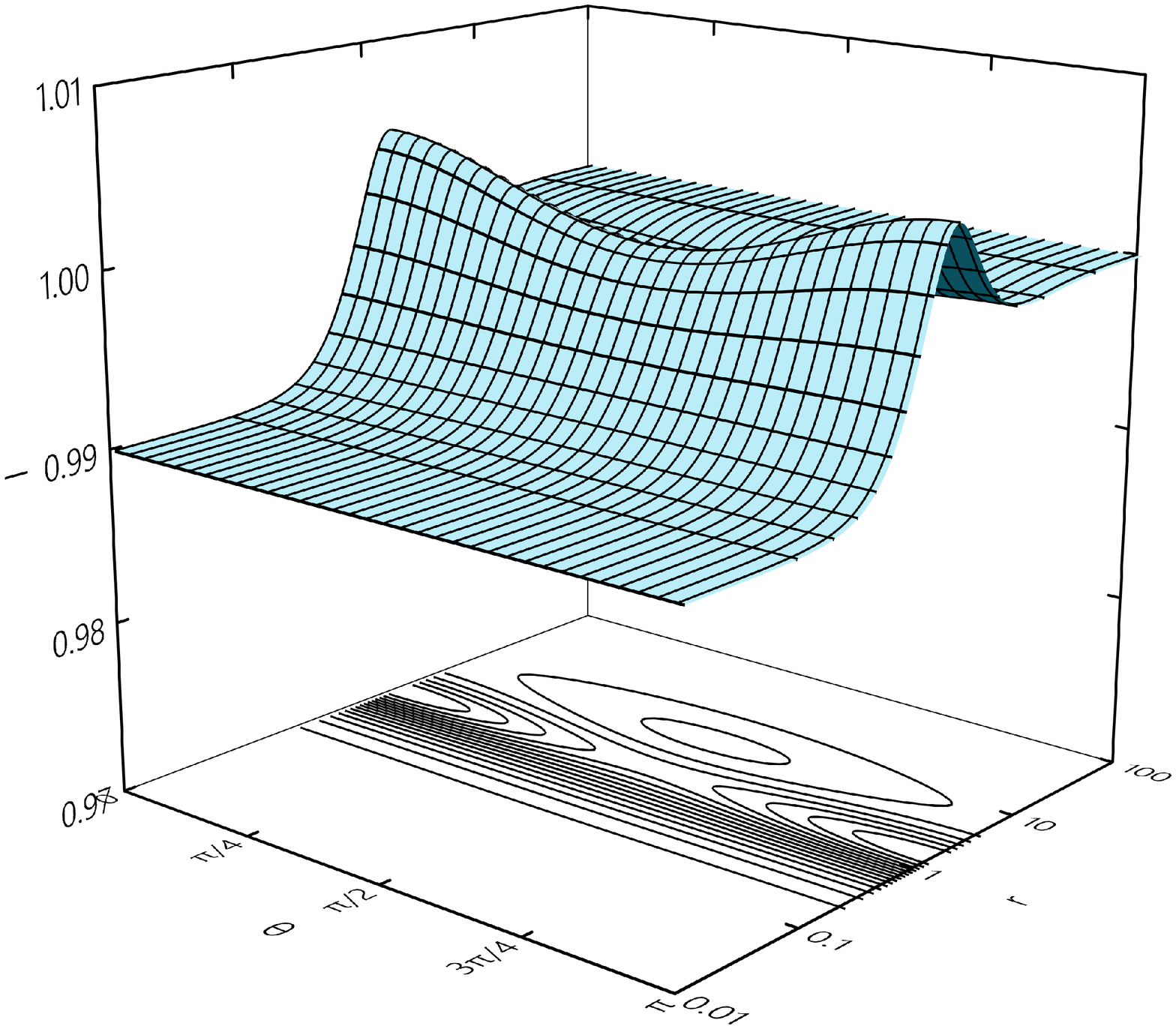}
\includegraphics[width=.49\textwidth, trim = 40 30 100 20, clip = true]{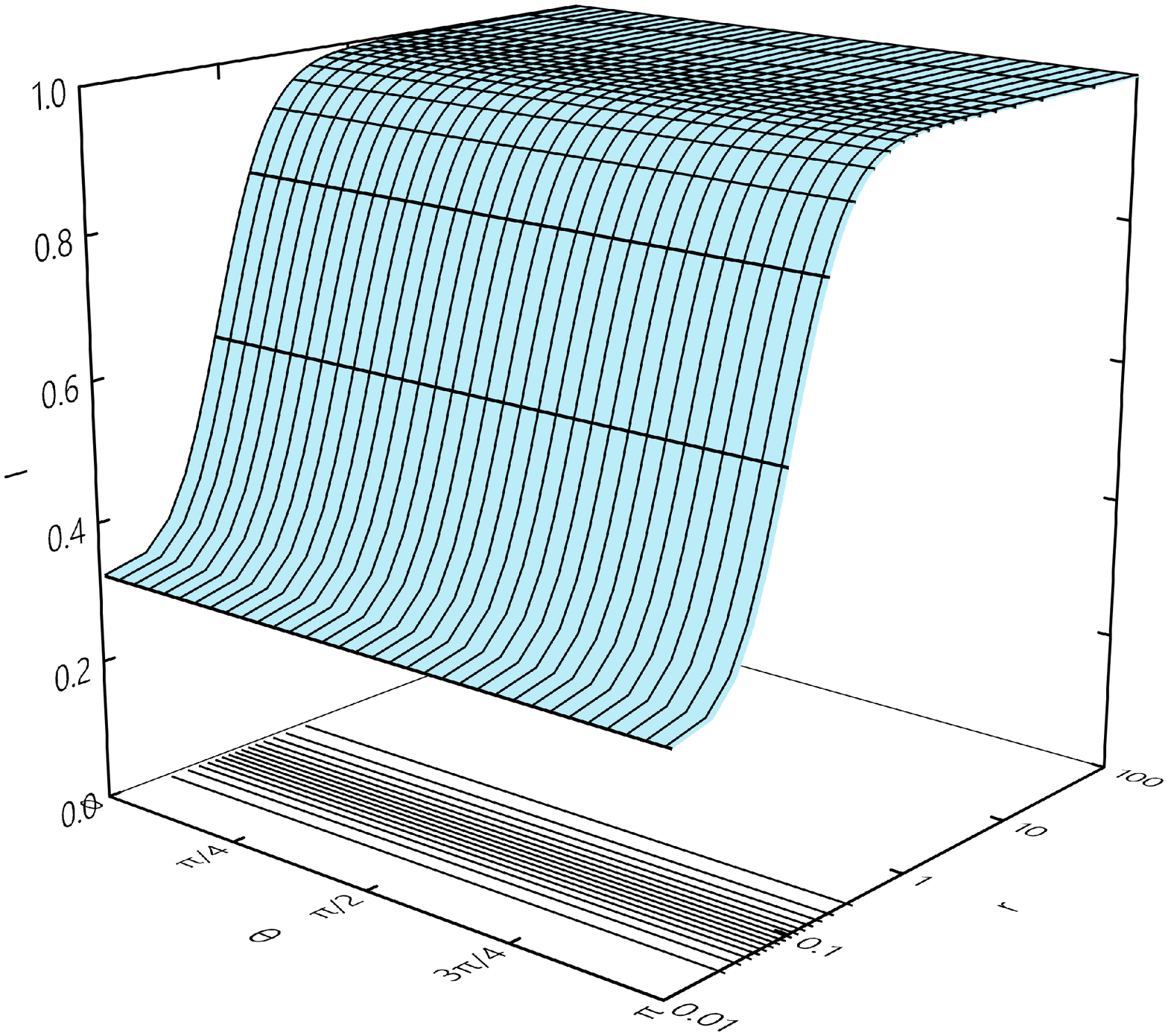}
\end{center}
\caption{\small
Distributions of metric function $l$ for `cloudy skyrmion' (left panel) and `cloudy BM' (right panel) solution with $c=0,\alpha=0.15, \omega=0.97$.
}
\end{figure}

\begin{figure}[hbt]
\lbfig{3do}
\begin{center}
\includegraphics[width=.49\textwidth, trim = 30 30 100 20, clip = true]{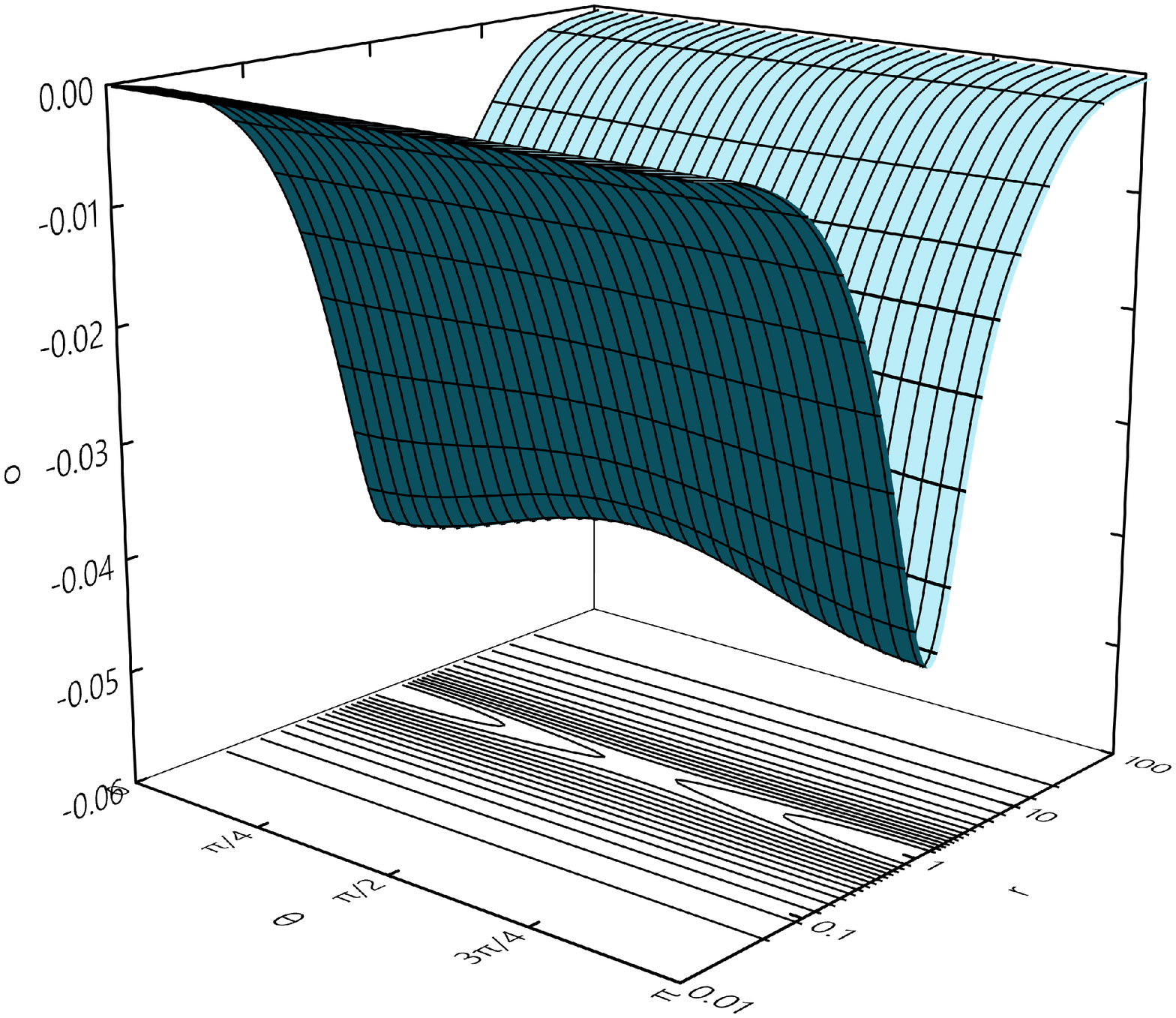}
\includegraphics[width=.49\textwidth, trim = 30 30 100 20, clip = true]{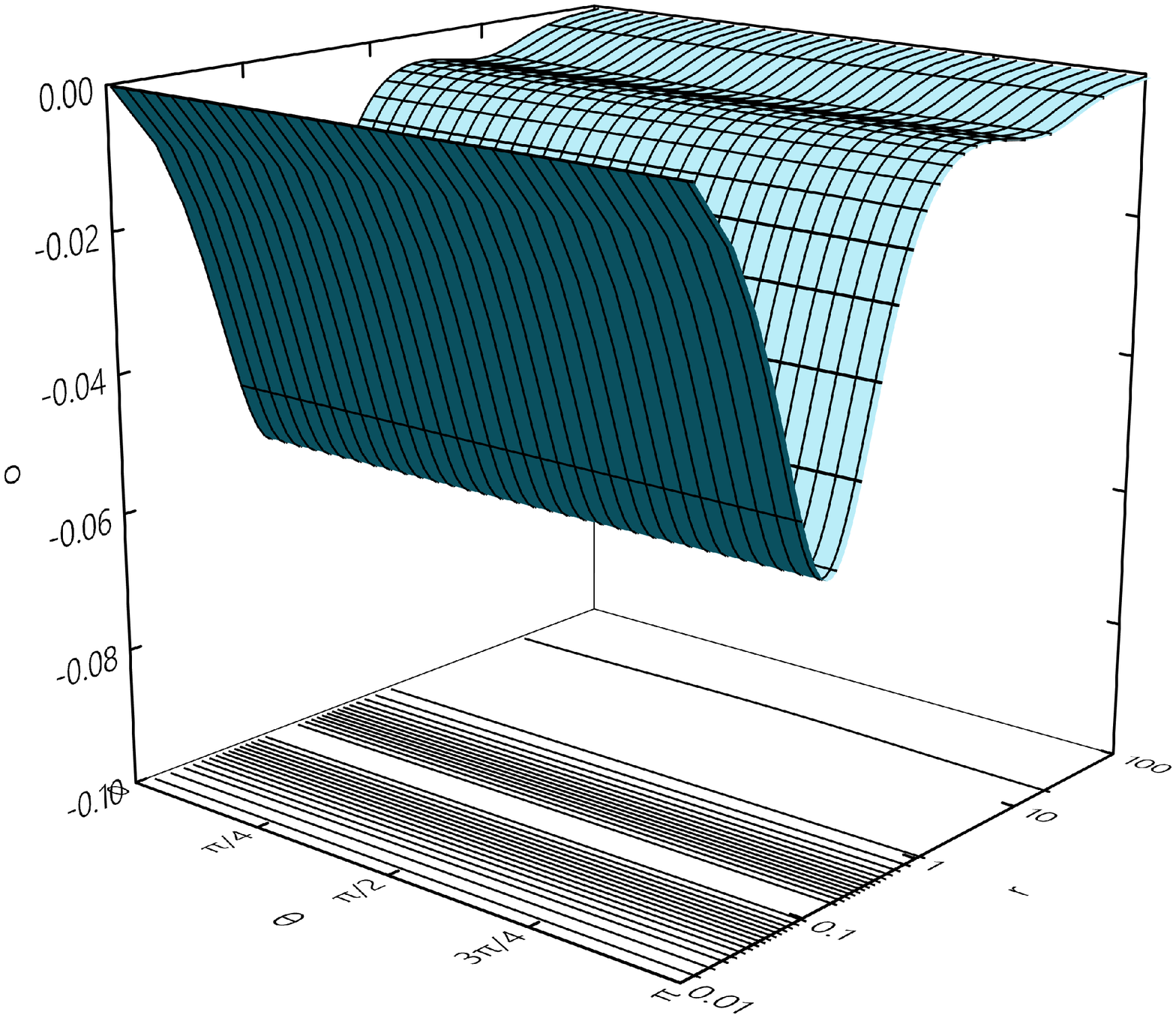}
\end{center}
\caption{\small
Distributions of metric function $o$ for `cloudy skyrmion' (left panel) and `cloudy BM' (right panel) solution with $c=0,\alpha=0.15, \omega=0.97$.
}
\end{figure}

\begin{figure}[hbt]
\begin{center}
\includegraphics[height=.28\textheight, trim = 60 20 80 50, clip = true]{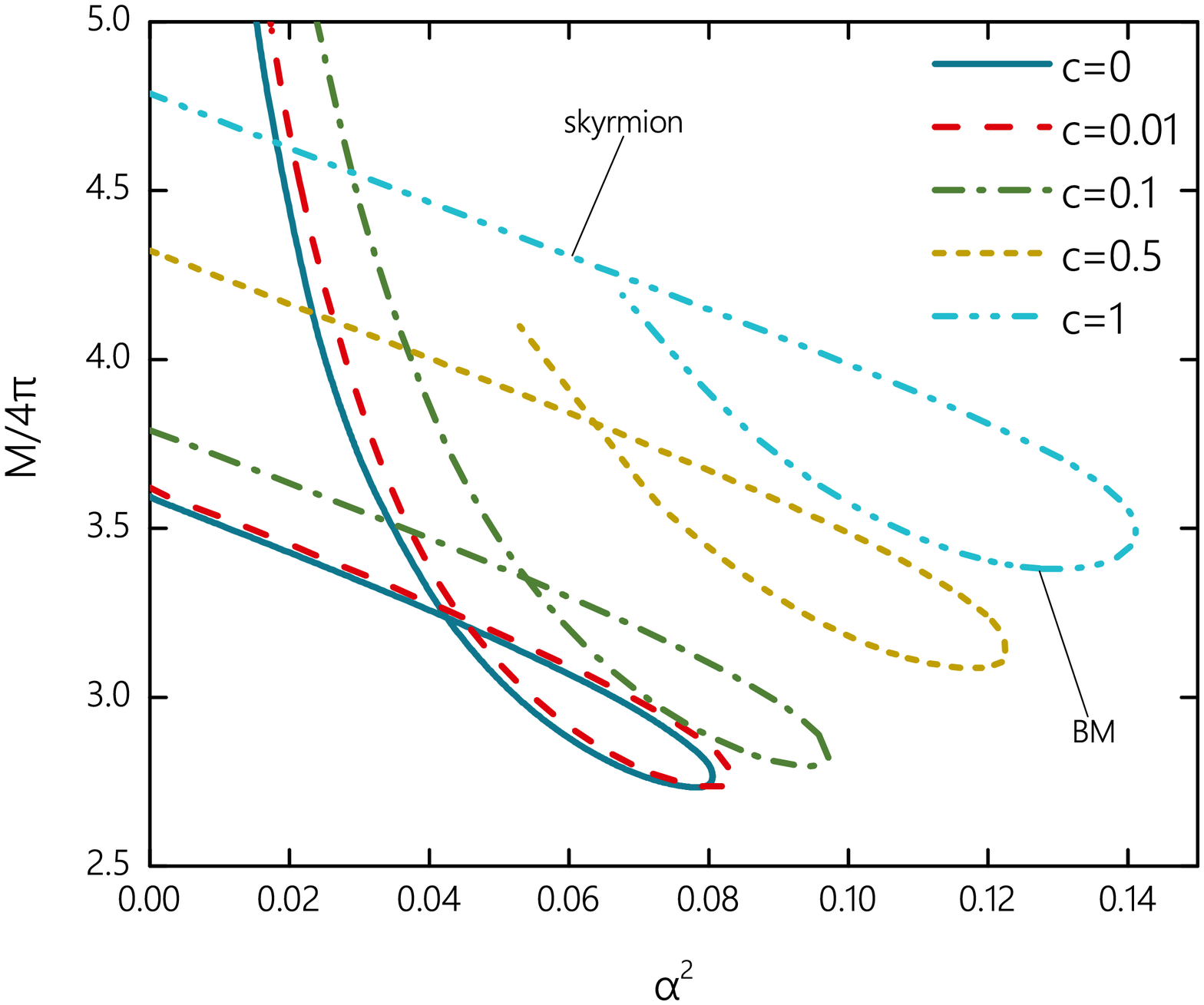}
\includegraphics[height=.28\textheight, trim = 60 20 80 50, clip = true]{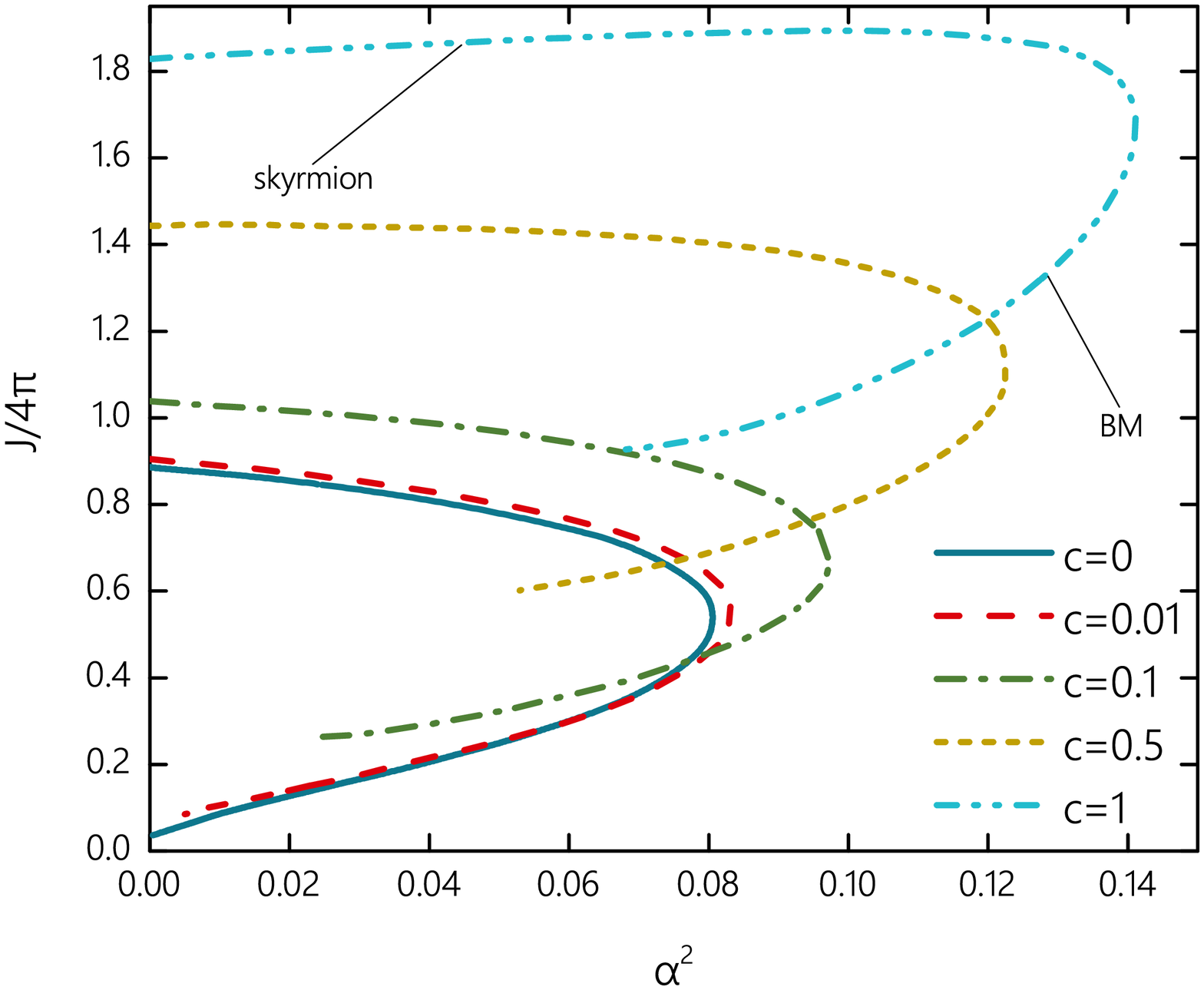}
\includegraphics[height=.28\textheight, trim = 60 20 80 50, clip = true]{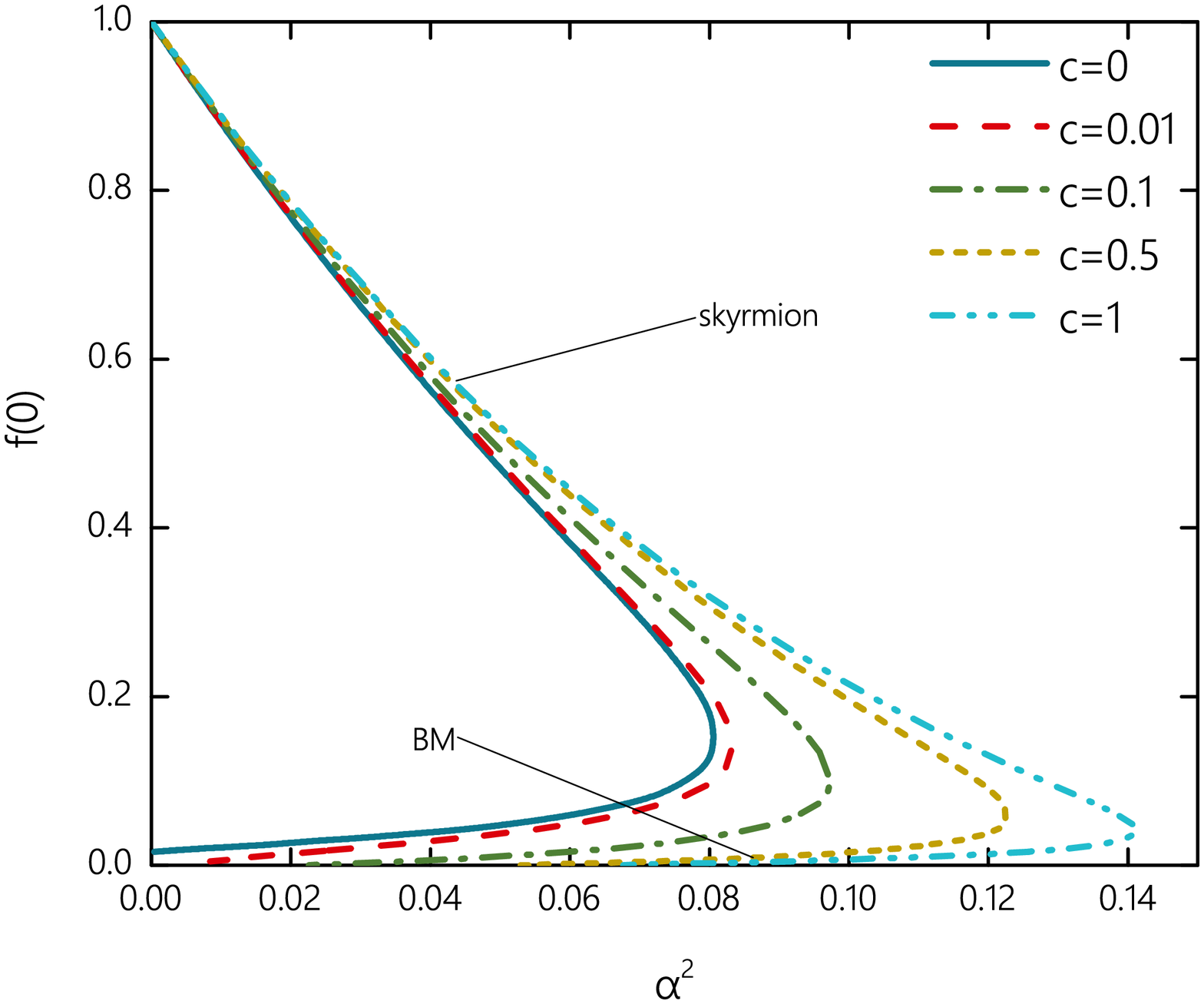}
\includegraphics[height=.28\textheight, trim = 60 20 80 50, clip = true]{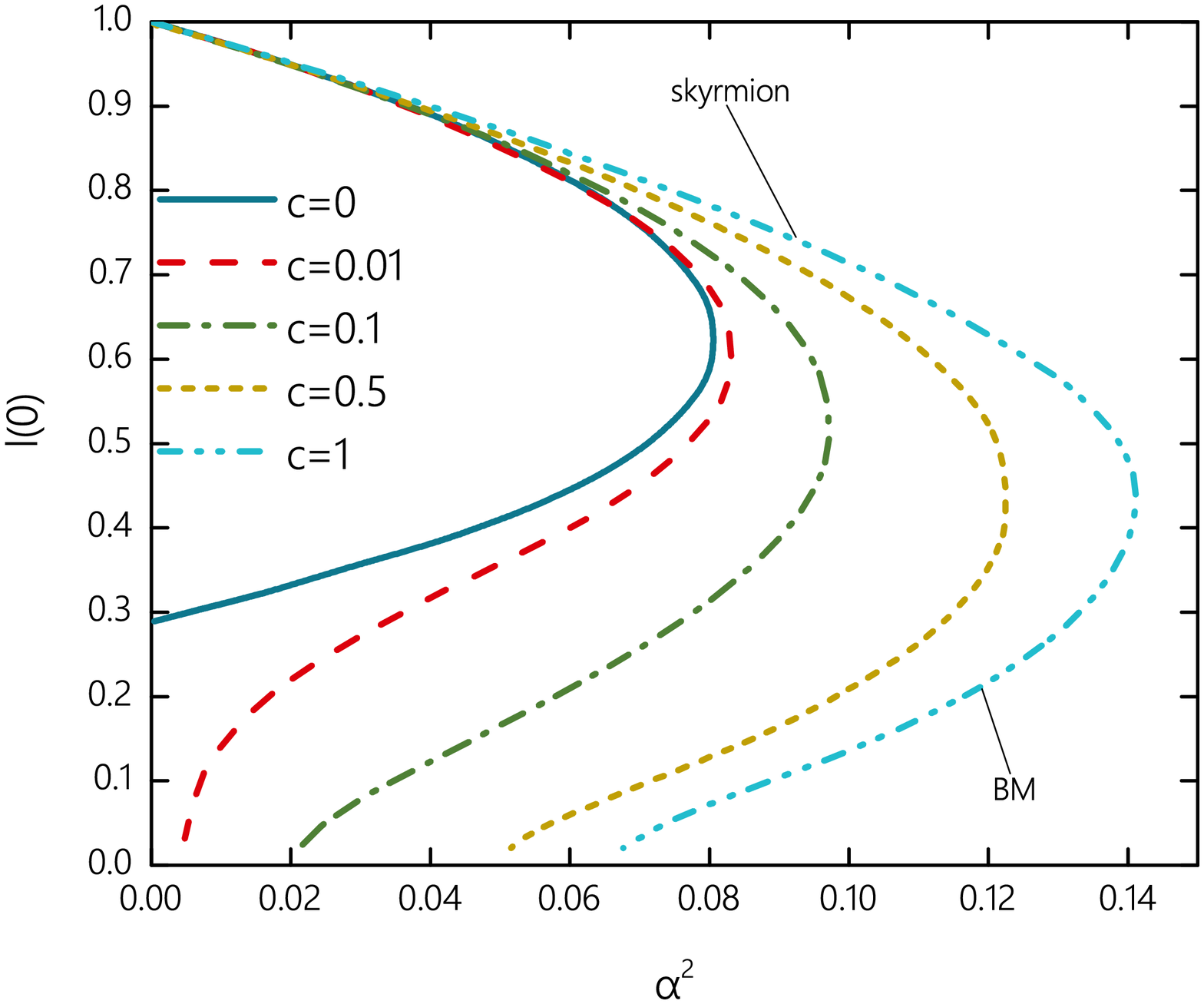}
\end{center}
\caption{\small
Dependencies of total mass $M$ (left upper plot), total angular momentum $J$
(right upper plot), value of metric function $f$ at the origin (left bottom plot) and
of value of metric function $l$ at the origin (right bottom plot) on
$\alpha^2$ for $\omega=0.7$ and different values of the sextic term
coupling constant $c$.
}
\lbfig{constomega}
\end{figure}

The `cloudy' branches appear independently, in the limit $\alpha \to 0$
they both bifurcate with the `pion cloud' solutions \cite{Ioannidou:2006nn}.
In the limiting case  ($c=0$) of the usual Einstein-Skyrme model,
as the angular frequency remains below the critical value
$\omega_{cr}=0.918$, the `cloudy Skyrmion' and `cloudy BM' branches
are disconnected both from the upper and lower branches of spinning gravitating
Skyrmions\footnote{This value is a bit lower than the corresponding critical frequency in the model with the usual
pion mass potential \cite{Ioannidou:2006nn}, the double-vacuum potential \re{doublevac} yields stronger
attraction.}. Instead, the `cloudy Skyrmion' branch merge with the `cloudy BM' branch at some second
critical value of the gravitational coupling, which is larger than $\alpha_{cr}$.

\begin{figure}[hbt]
\begin{center}
\includegraphics[height=.28\textheight, trim = 50 20 80 50, clip = true]{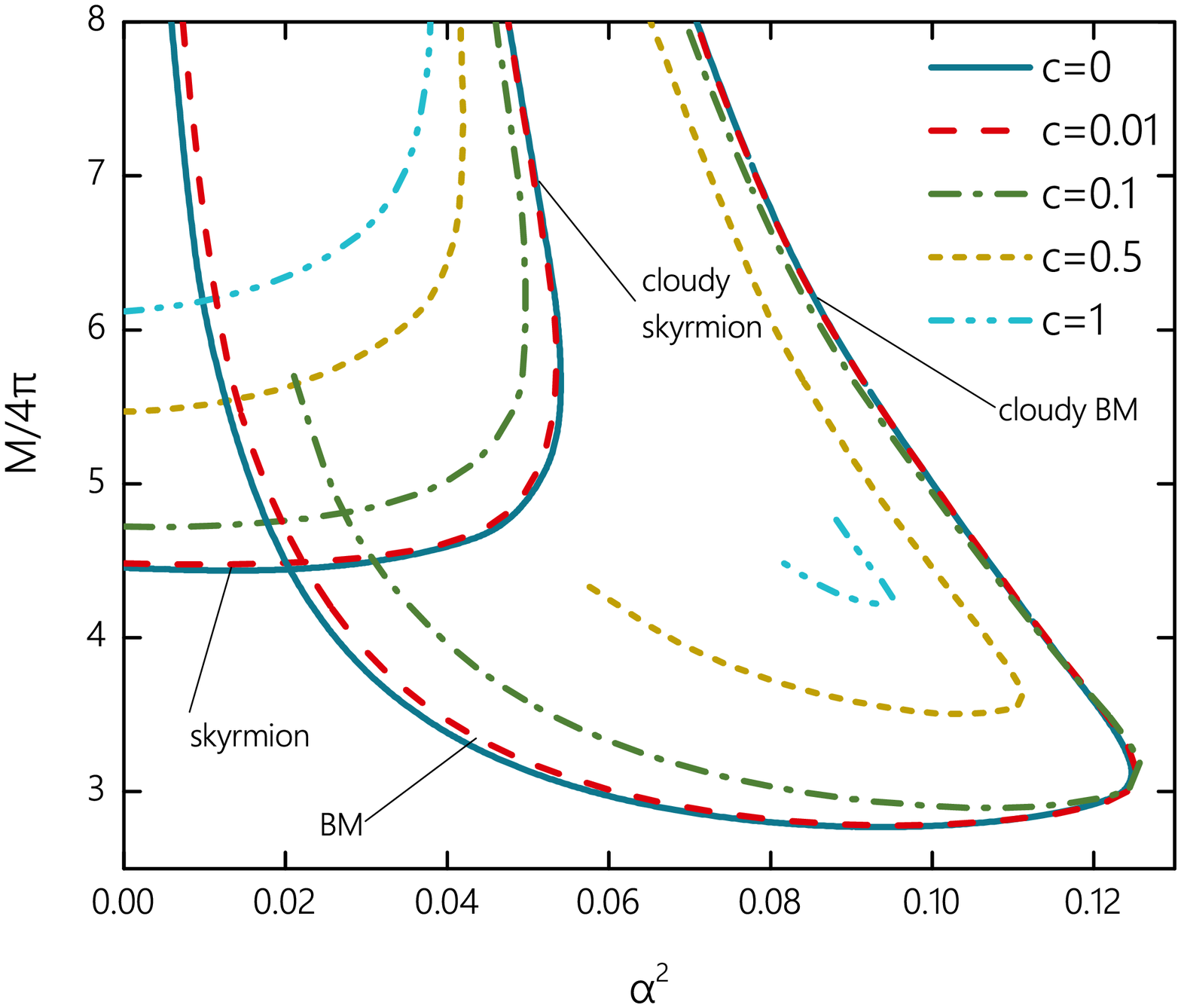}
\includegraphics[height=.28\textheight, trim = 50 20 80 50, clip = true]{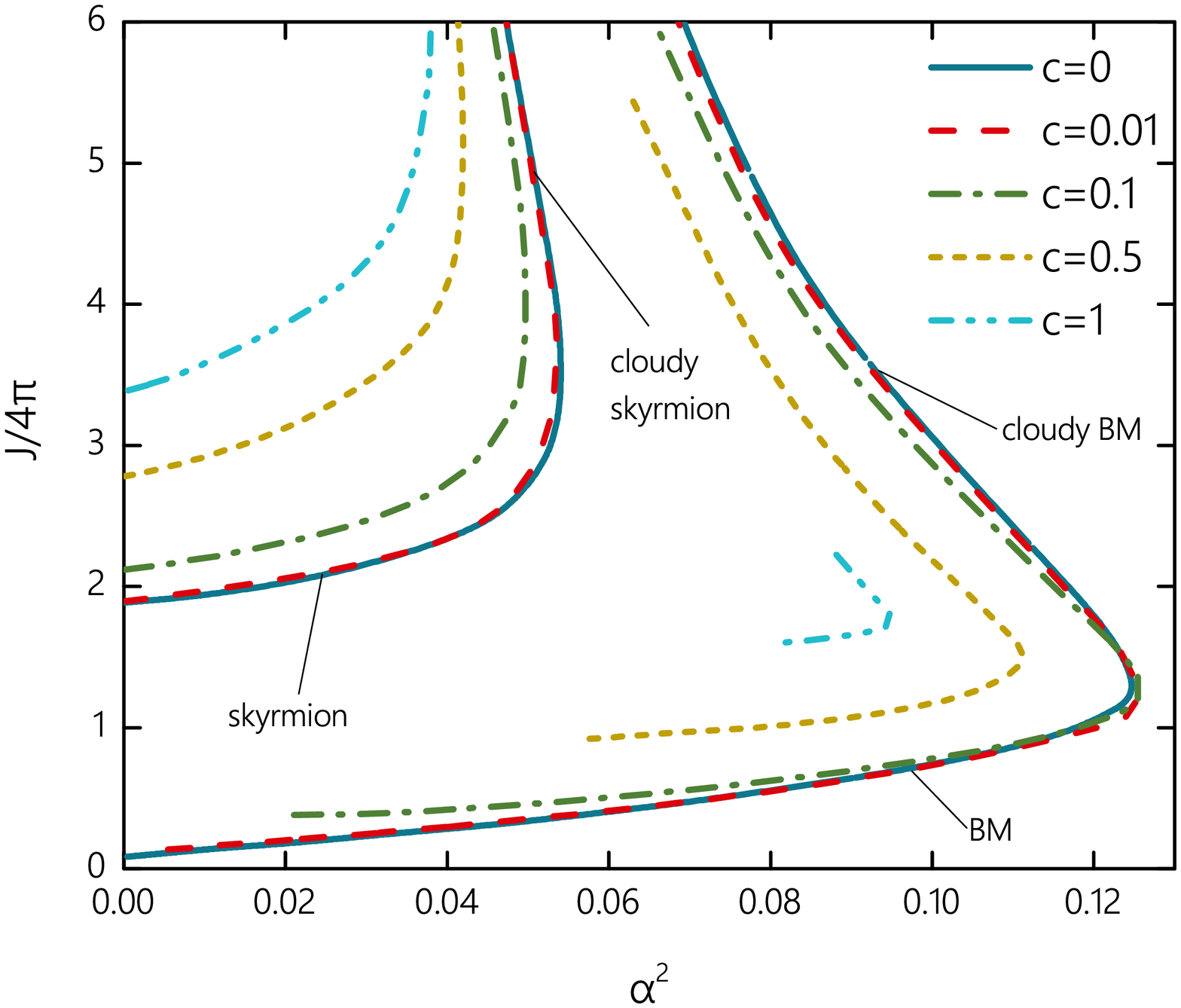}
\includegraphics[height=.28\textheight, trim = 50 20 80 50, clip = true]{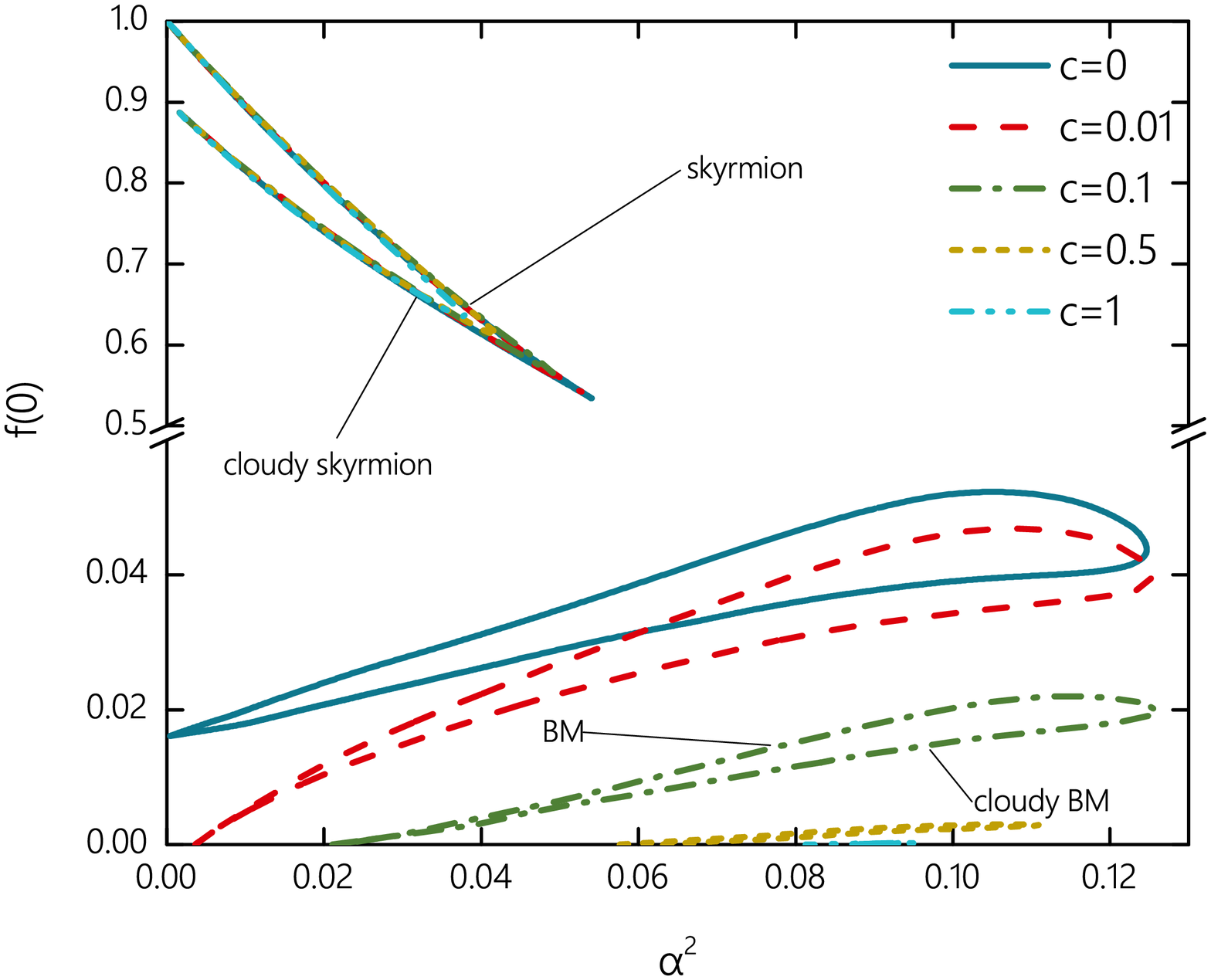}
\includegraphics[height=.28\textheight, trim = 50 20 80 50, clip = true]{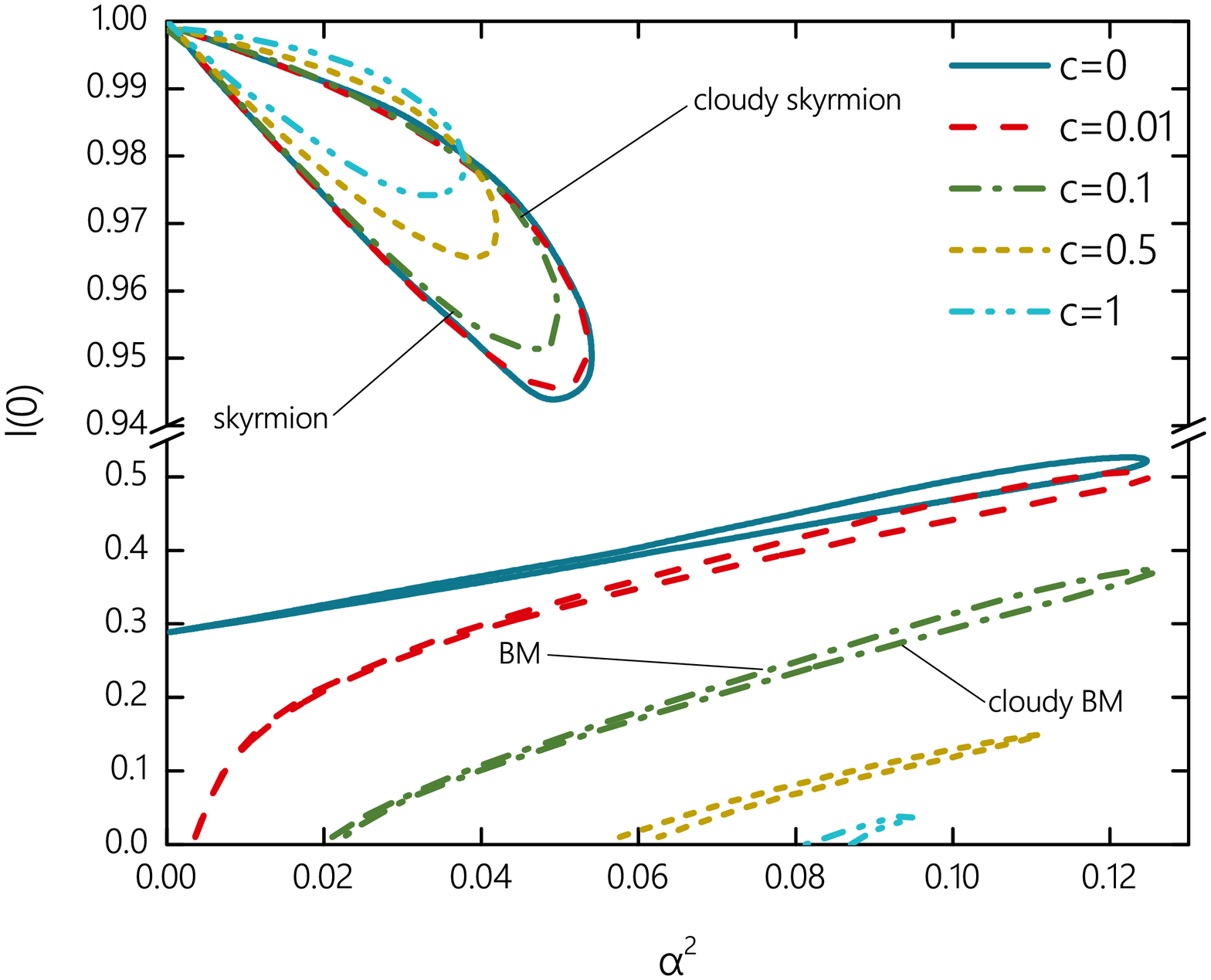}
\end{center}
\caption{\small
Dependencies of total mass $M$ (left upper plot), total angular momentum $J$
(right upper plot), value of metric function $f$ at the origin (left bottom plot) and
of value of metric function $l$ at the origin (right bottom plot) on $\alpha^2$
for $\omega=0.97$ and different values of sextic term
coupling constant $c$.
}
\lbfig{constomega-2}
\end{figure}

However, the situation changes as the frequency $\omega$ increases above this threshold.
Considering the interplay between the
spinning Skyrmions and cloudy configurations in the general Einstein-Skyrme model, we observe that,
similar to the case of the model without the sextic term \cite{Ioannidou:2006nn},
for $\omega > \omega_{cr}$ and any non zero value of the coupling $c$,
the  `Skyrmion' branch merges the `cloudy Skyrmion' branch, which extends all the way back to
the limit $\alpha=0$, see Figs.~\ref{constomega-2},\ref{angvel}.
The mass of the solutions on this branch is higher,
they can be viewed as bound states
of the gravitating spinning Skyrmion and pion excitations. Along this branch the coupling to the quadratic term
$a$ is decreasing,
similar to the case of the bifurcation between the
`Skyrmion' and `BM' branches above. The limiting $a=0$ configuration for any $c\neq 0$ corresponds to the
spinning compacton solutions of the
$\mathcal{L}_4+\mathcal{L}_6+\mathcal{L}_0$ submodel, in this limit the pion excitations are coupled to the
compacton by gravitational interaction.

As we noticed above, in the general Einstein-Skyrme model with non-zero contribution of the sextic term,
the limiting configuration becomes singular at some minimal value of the gravitational coupling
$\alpha_{min}$, it does not approach the rescaled BM solution, unless $c=0$.
However, we observe that the bifurcation into another
two $\alpha$-branches continues to hold, in the limit $\alpha\to \alpha_{min}$ they both are linked to
the corresponding singular solution, see Fig.~\ref{constomega-2}. The spinning configurations on the lower in energy
branch are counterparts of  the usual solutions on the `BM' branch. It
extends up to second maximal value of the gravitational
coupling, which is typically higher than the critical coupling,
at which the `Skyrmion' and `BM' branches bifurcate. There it merges with the `cloudy' branch, which
evolves backwards to the singular solution, the configurations on that branch are composite states of the spinning
soliton and pion excitations. As the coupling to the sextic term increases, these `singular' branches become shorter, see
Fig.~\ref{constomega-2}.

\begin{figure}[hbt]
\begin{center}
\includegraphics[height=.28\textheight, trim = 60 20 80 50, clip = true]{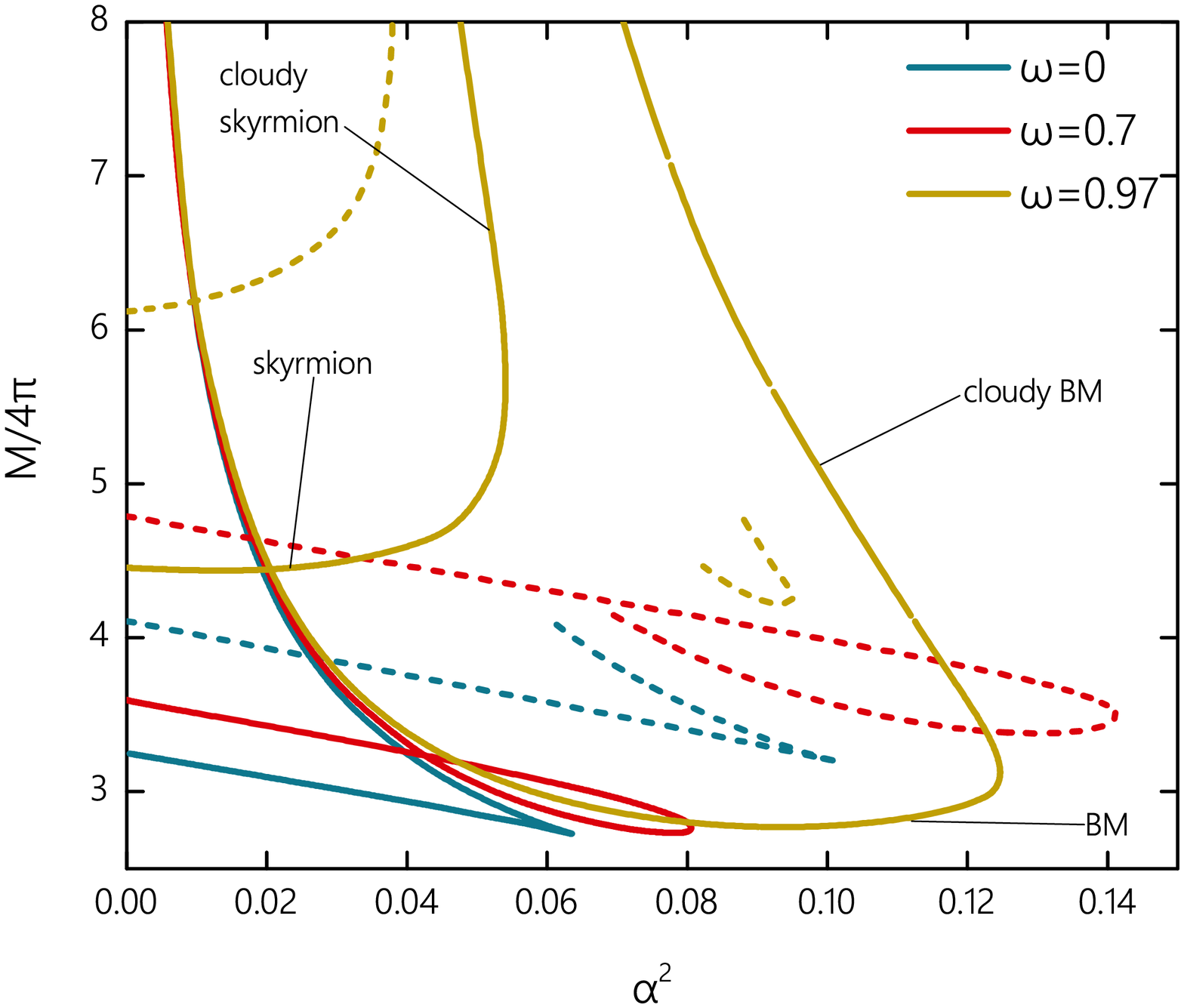}
\includegraphics[height=.28\textheight, trim = 60 20 80 50, clip = true]{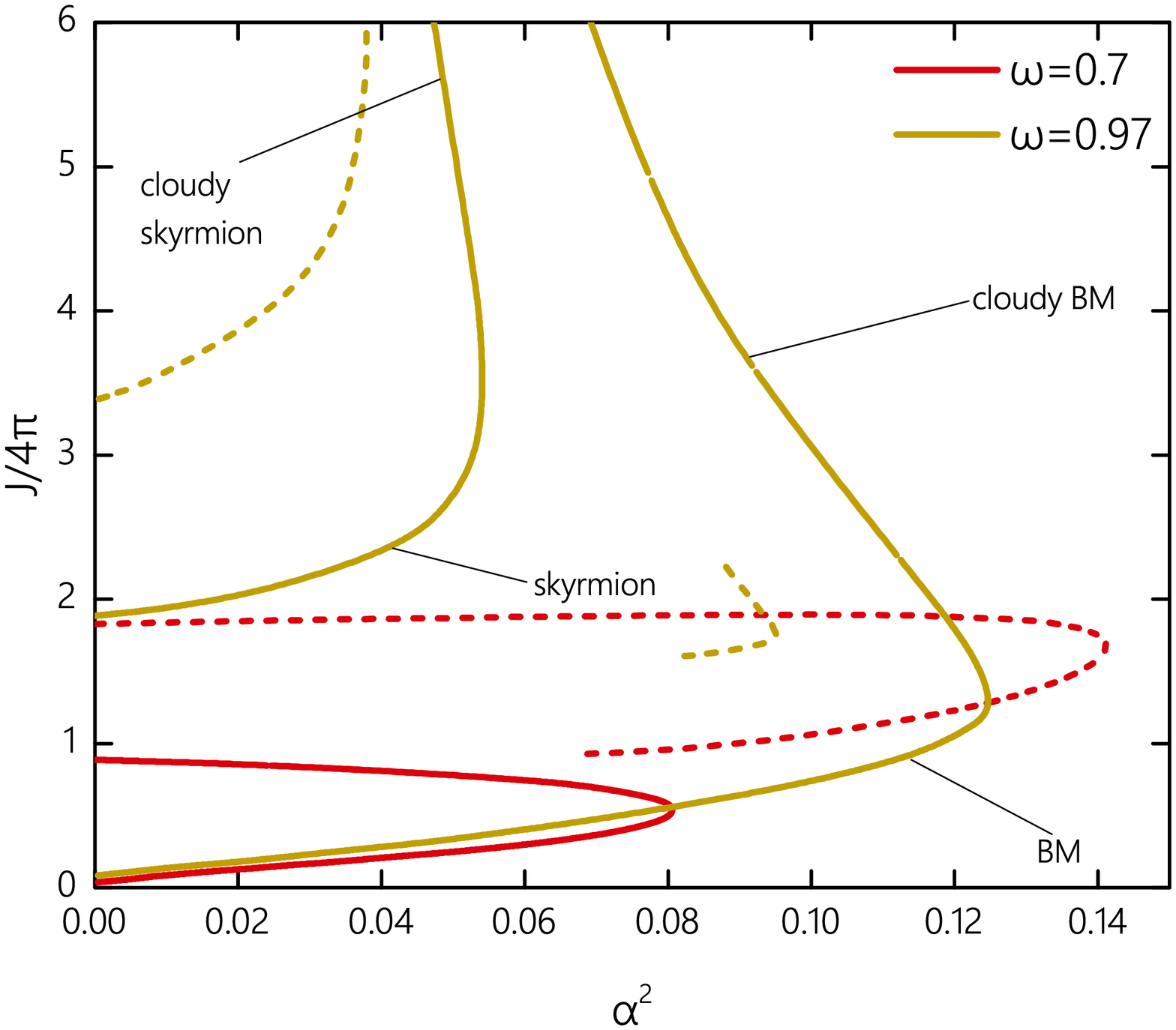}
\includegraphics[height=.28\textheight, trim = 60 20 80 50, clip = true]{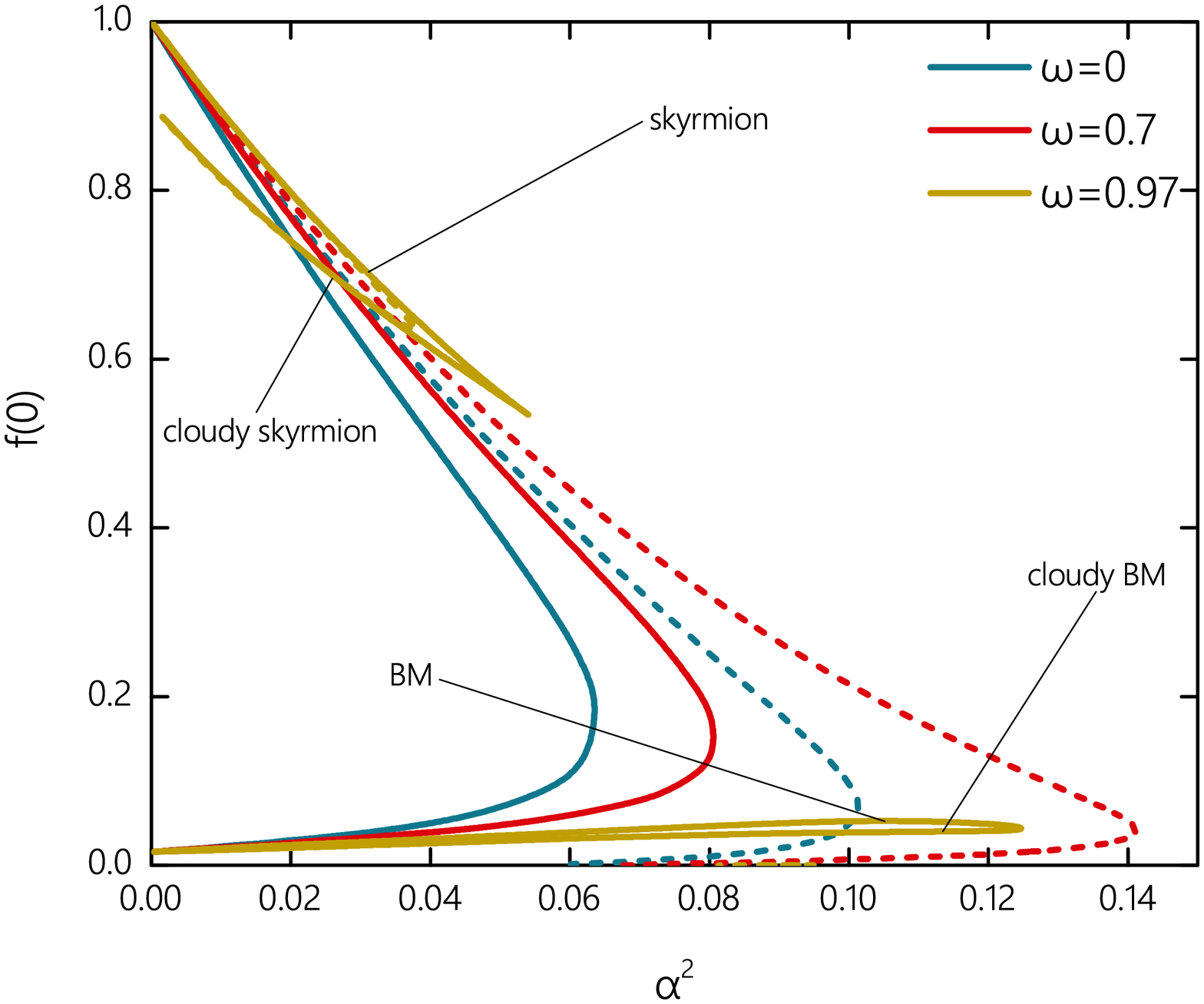}
\includegraphics[height=.28\textheight, trim = 60 20 80 50, clip = true]{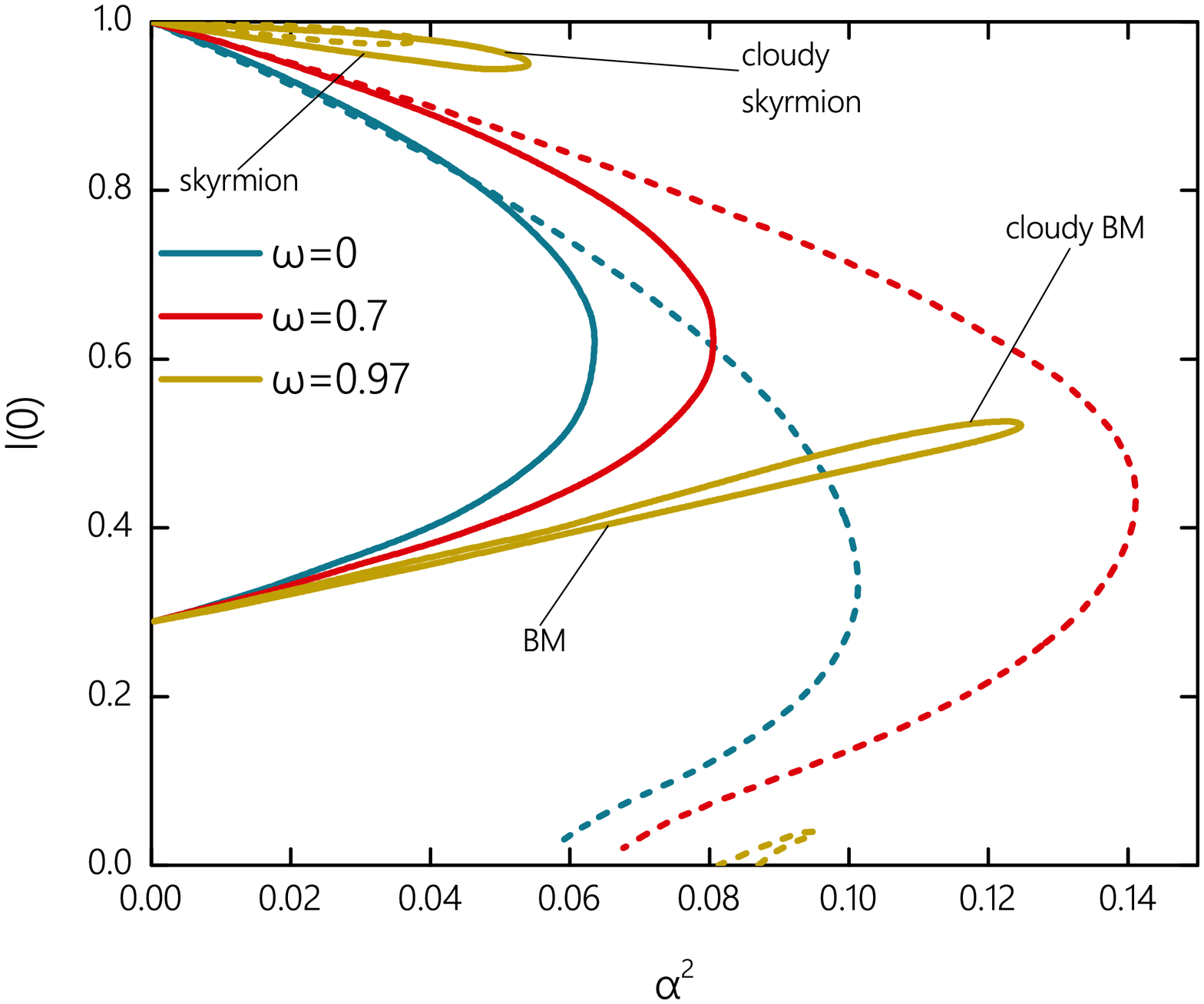}
\end{center}
\caption{\small
Dependencies of the mass of the solutions $M$ (left upper panel), the angular momentum $J$ (right upper panel),
the value of metric function $f$ at the origin (left lower panel) and the value of metric function $l$ at
the origin (right lower panel) are plotted as functions of the gravitational coupling $\alpha^2$  for  $c=0$
(solid lines) and $c=1$ (dashed lines), and some set of values of the angular  frequency $\omega$.
}
\lbfig{angvel}
\end{figure}

Numerical results show that on the `BM' branch and its counterpart
the angular momentum $J$ and the values of the metric functions at the origin $f(0)$, $l(0)$
decrease monotonically as $\alpha$ decreases. On the other hand, the mass of the solutions
possess a minimum on this branch, then it starts to increase approaching the mass of the rescaled
limiting solution, see Fig.~\ref{angvel}.

The value of the threshold frequency, $\omega_{cr}$, also depends on the value of $c$ it
decreases as $c$ increases. In particular, for general Einstein-Skyrme model with $c=1$
the value of threshold frequency is $\omega_{cr}\sim 0.76$. This dependency leads to a bit more
complicated pattern of branch structure than we described above. Since the pion clouds, as well as the
cloudy branches, exist only for relatively large values of angular frequency, in some range of frequencies
$\omega>\omega_{cr}$, both the `Skyrmion' branch and `BM' branch just terminate at some values of the
gravitational parameter $\alpha$.

\begin{figure}[hbt]
\begin{center}
\includegraphics[height=.28\textheight, trim = 60 20 80 50, clip = true]{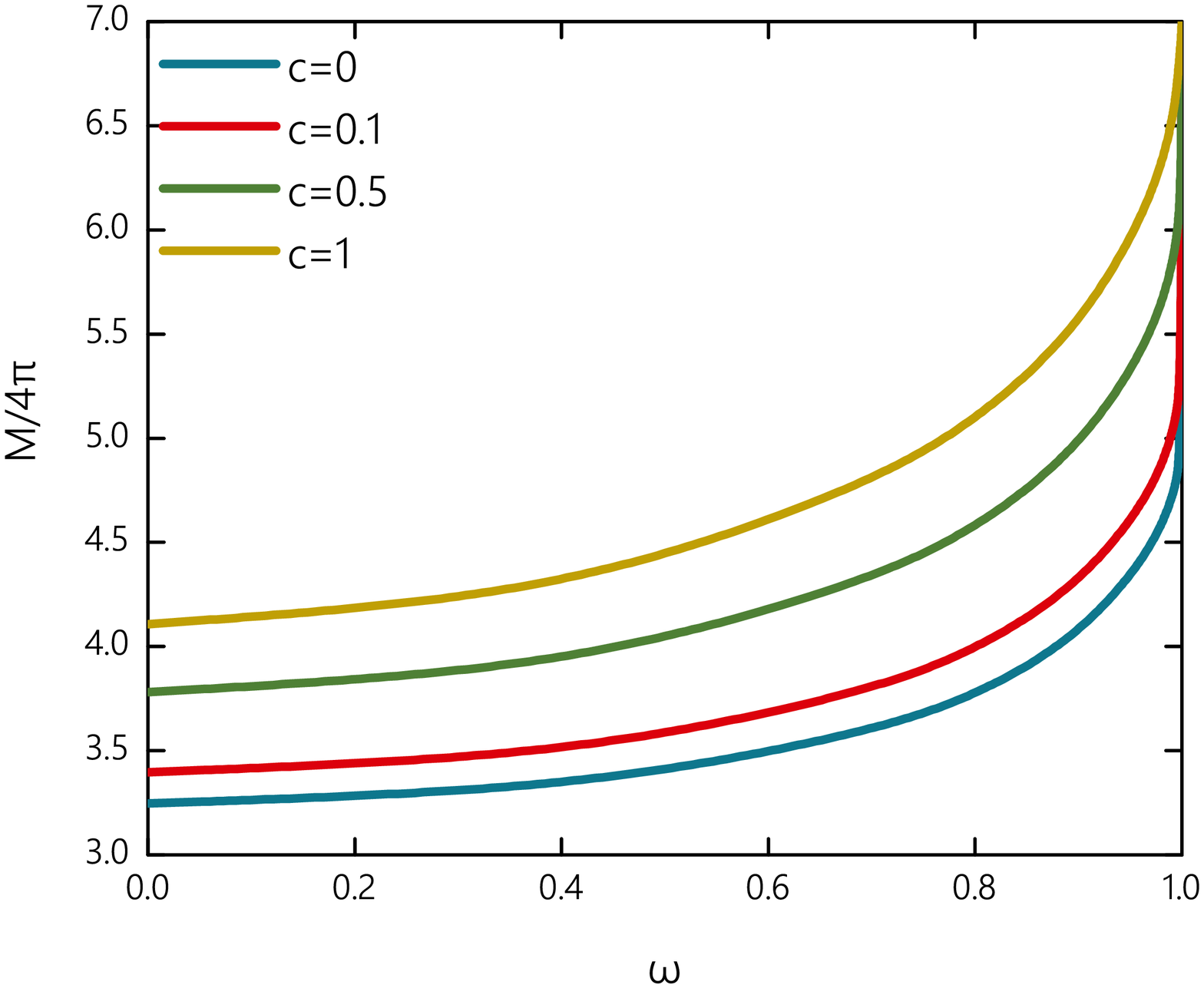}
\includegraphics[height=.28\textheight, trim = 60 20 80 50, clip = true]{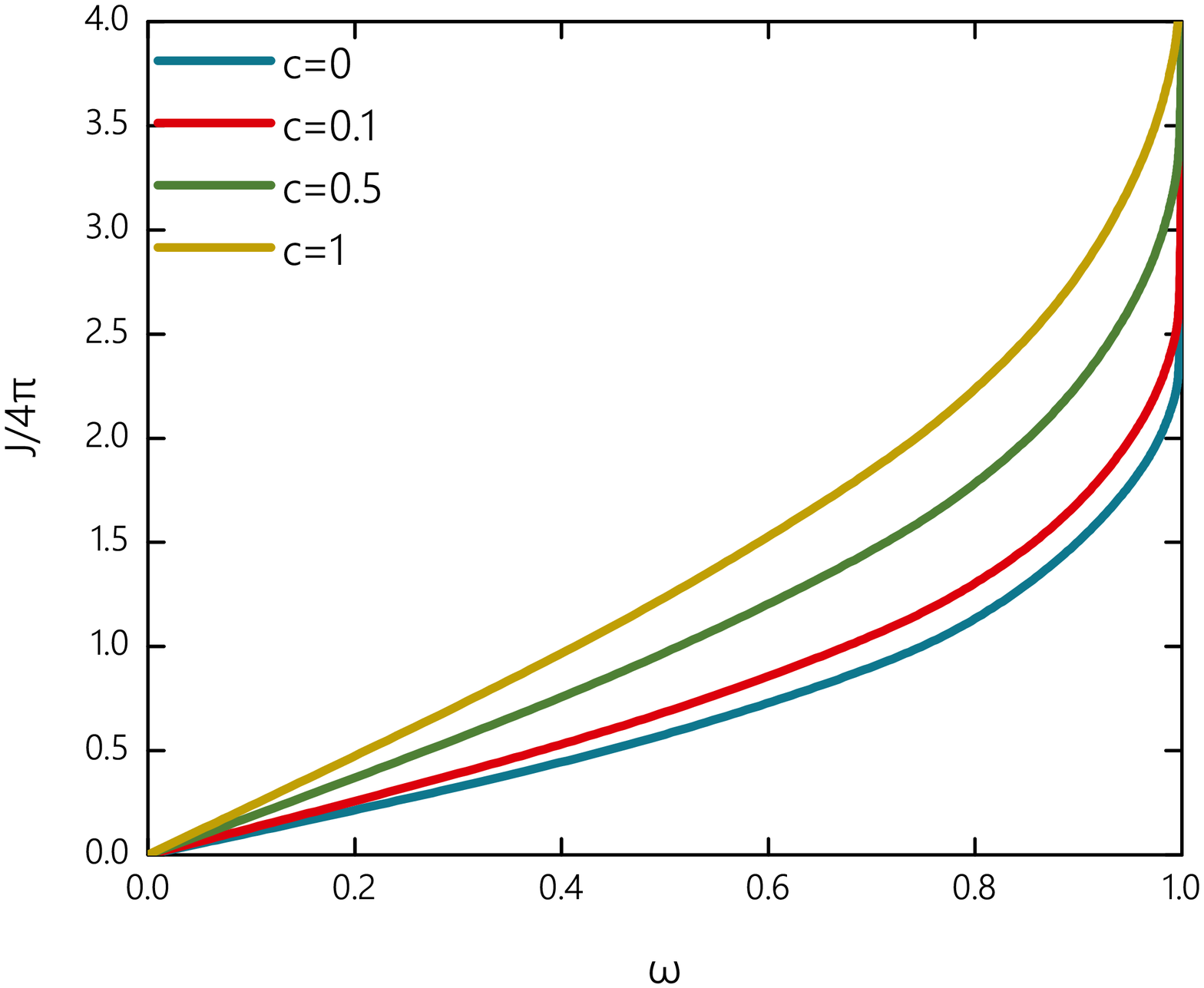}
\end{center}
\caption{\small
The  mass $M$ (left upanel) and the angular momentum $J$ (right panel) of spinning solutions are plotted as
functions of the angular frequency $\omega$ for $\alpha=0.01$
and a few values of the coupling constant $c$.
}
\lbfig{MJomega}
\end{figure}

In  Fig.~\ref{MJomega} we display dependencies of the mass $M$ and the angular
momentum $J$ of the `Skyrmion' branch solutions on the
angular frequency for some fixed value of the gravitational coupling $\alpha$.  As expected, we observe
that both the  mass and the angular momentum increase both with increasing $\omega$, and the coupling constant $c$.
Notably, for the wide range of values of the angular frequency the angular momentum grows almost linearly as
$\omega$ increases. This is in agrement with corresponding observations in the case of
spinning Skyrmions in the flat space \cite{Battye:2014qva,Halavanau:2013vsa,Battye:2013tka}.
As $\omega$ approaches the upper critical value, the angular
momentum rapidly increases for all values of the coupling constant $c$.

\begin{figure}[hbt]
\begin{center}
\includegraphics[height=.28\textheight, trim = 60 20 80 50, clip = true]{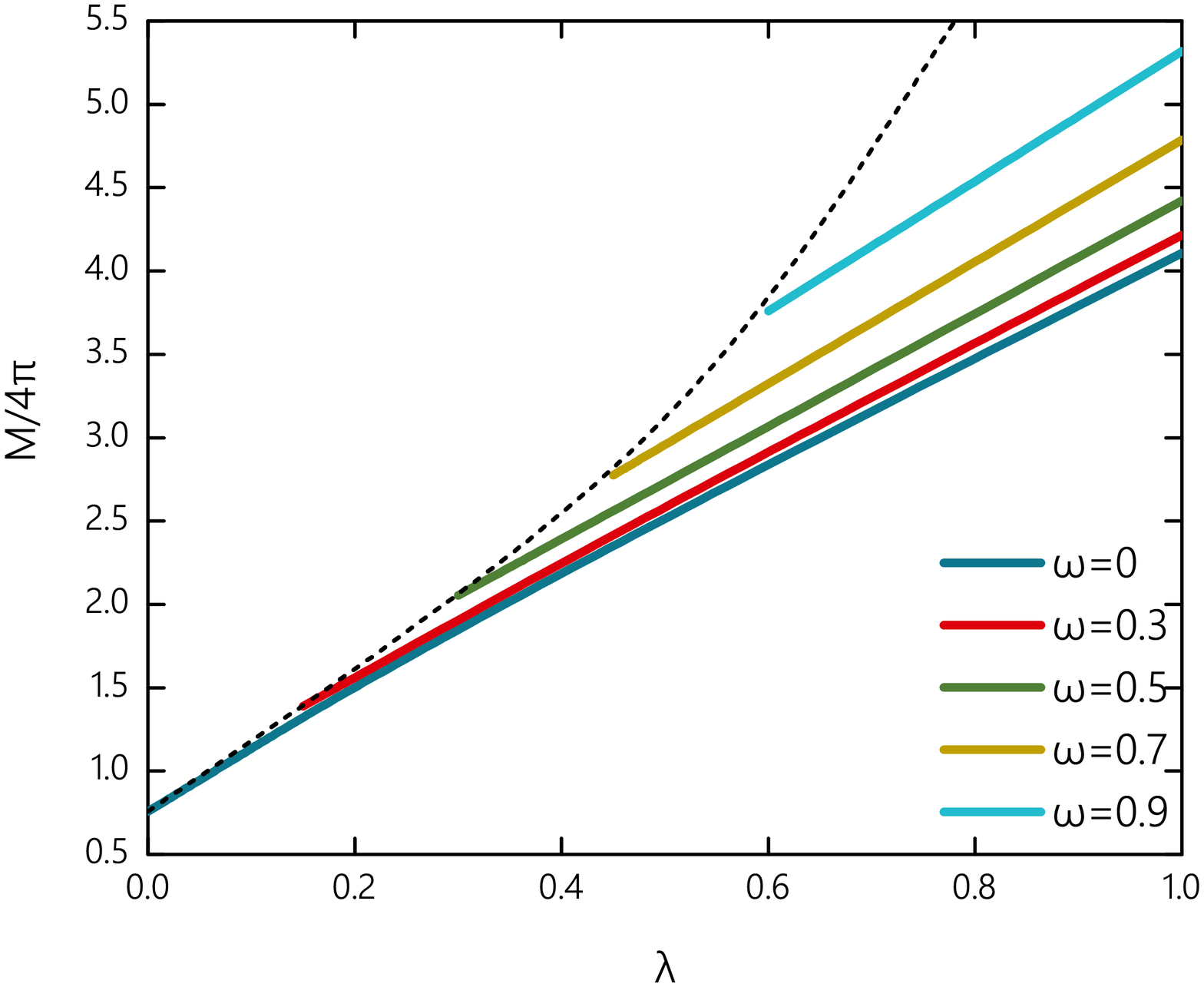}
\includegraphics[height=.28\textheight, trim = 60 20 80 50, clip = true]{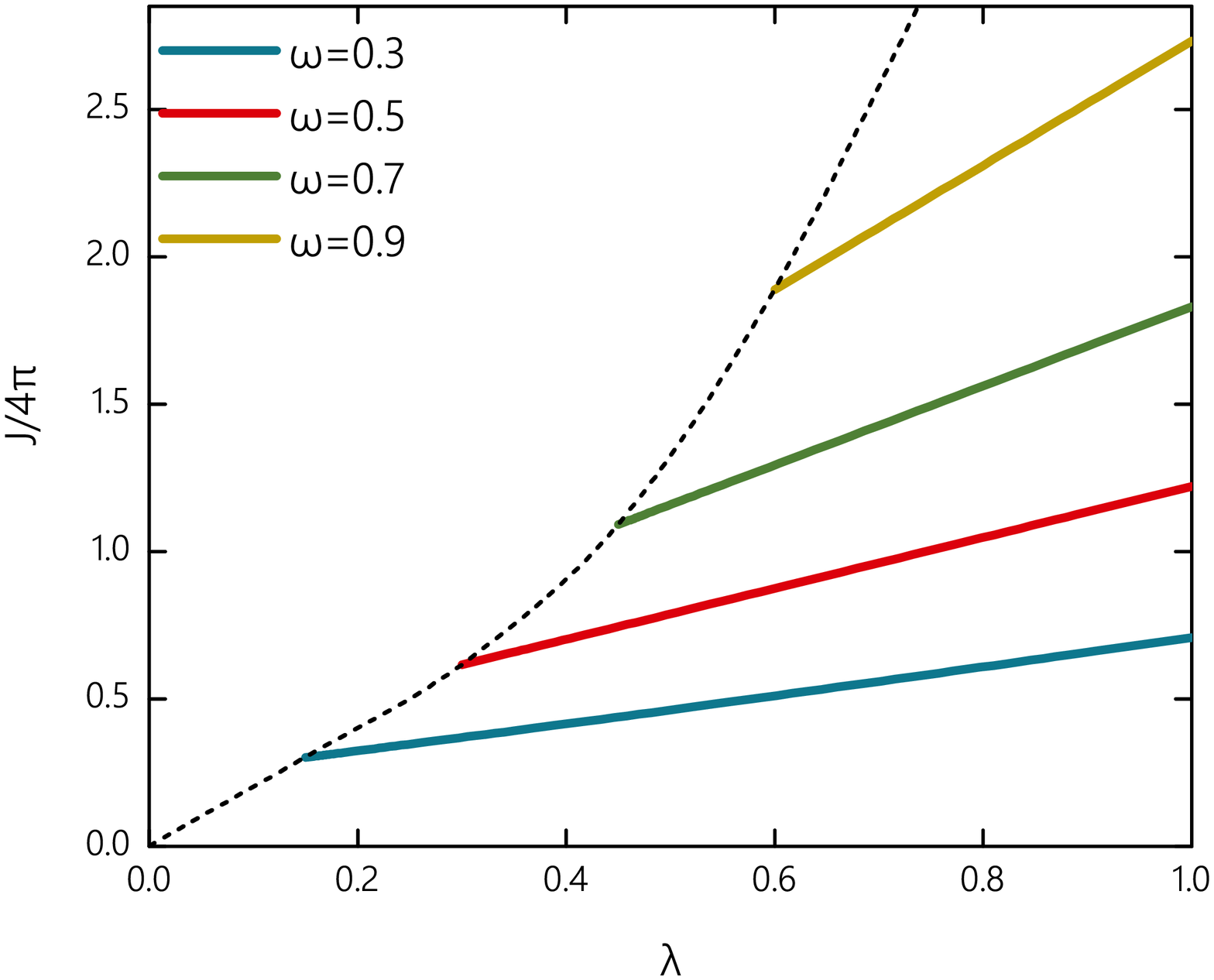}
\end{center}
\caption{\small
Dependencies of the mass $M$ (left plot) and the angular momentum $J$ (right plot)
on the value of the parameter $\lambda=a=b$ for $\alpha=0.01,~ c=1$
and some set of values of the angular frequency  $\omega$.
Dashed line indicates end-points, at which the corresponding solution cease to exist.
}
\lbfig{BPS}
\end{figure}

Finally, let us consider the situation when the general Einstein-Skyrme model is approaching the limiting
$\mathcal{L}_6+\mathcal{L}_0$ self-dual submodel. We fix the coupling
$c=1$ and
simultaneously decrease both parameters $a=b=\lambda$ considering different values of the angular frequency
$\omega$. We observe, that, as $\lambda$ decreases, both the mass and angular momentum of the spinning
solitons decrease almost linearly, see Fig.~\ref{BPS}. However, at some critical value of the
parameter $\lambda_{cr}$ the metric functions approach singular limit and numerical errors
increase dramatically. We can expect that in this limit both
the ansatz for the metric
\re{metric} and the axially symmetric parametrization of  the Skyrme field \re{field} are no longer applicable.
Physically it may be related with decay of the configuration which becomes torn apart.

The value of $\lambda_{cr}$ rapidly
decreases as $\omega$ increases, from $\lambda_{cr}=0$ at $\omega=0$, it slightly depends on
the gravitational coupling $\alpha$, but nevertheless it remains finite  as $\alpha$ tends to zero.
Thus, we can conclude that there are no spinning solitons neither in the reduced
$\mathcal{L}_6+\mathcal{L}_0$ Einstein-Skyrme submodel, nor in the flat space.

\section{Summary and conclusions}
The main purpose of this work was to construct globally regular axially symmetric
 stationary rotating solutions of the generalized
Einstein-Skyrme in asymptotically flat spacetime. We study the dependence of the field configurations
on the gravitational coupling parameter $\alpha$ and on the angular frequency $\omega$.
There is a complicated pattern of branches for fixed angular frequency, which, however always includes
a lower branch of gravitating $\alpha$-branch of spinning solutions.
This branch emerges from the corresponding
flat-space configuration in the limit $\alpha \to 0$.
For small-to-moderate values of
the angular frequency $\omega < \omega_{cr}$,
both in the generalized  model with the sextic term in the matter field
sector, and in the conventional Einstein-Skyrme model,
this branch merges another branch of regular solutions at some maximal value of gravitational coupling.
While for the usual $\mathcal{L}_2+\mathcal{L}_4+\mathcal{L}_0$ submodel it extends back to the limit of vanishing gravitational
coupling, where it tends to the corresponding rescaled BM solution, in the generalized theory
with sextic term, the upper branch of solutions approaches singular solution at finite minimal value of $\alpha$.
Further, as the angular frequency increases above some critical value $\omega_{cr} \lesssim \omega_{max}$, two different
`cloudy' branches appear, they represent bound states of spinning Skyrmions and topologically trivial excitation, the
`pion clouds' \cite{Ioannidou:2006nn}. Then the `cloudy skyrmion' branch merges the Skyrme branch of solutions
while the `cloudy BM' branch merges the BM branch. In the generalized model, for any $c\neq 0$,
the evolution along the second "cloudy skyrmion"  branch exists all the way back to the
limiting $\alpha=0$ configuration, which represents a
spinning compacton solutions of the
$\mathcal{L}_4+\mathcal{L}_6+\mathcal{L}_0$ submodel coupled to the `pion'  excitations by the
strong gravitational attraction. On the other hand, the upper branch of solutions in such a case, bifurcates
with a counterpart of the `cloudy BM' branch, they both terminates at the corresponding singular solutions at some
finite values of the gravitational coupling.

An important observation is that the truncated $\mathcal{L}_6+\mathcal{L}_0$ Einstein-Skyrme submodel
does not support regular stationary spinning Skyrmions, they do not exist both in flat and curved space.

In this paper, we have considered Skyrmion solution only of topological degree one,
As a direction for future work, it would be interesting
to study the higher charge spinning gravitating Skyrmions. Another interesting question, which we hope to
be addressing in the near future, is
to construct spinning generalizations of the black holes with Skyrmionic hair, which would provide
a 3+1 dimensional counterpart of the 4+1 dimensional configurations presented recently in \cite{Brihaye:2017wqa}.

\section*{Acknowledgements}
We are grateful to Burkhard Kleihaus and Jutta Kunz for inspiring and valuable discussions
Y.S. gratefully
acknowledges support from the Russian Foundation for Basic Research
(Grant No. 16-52-12012), the Ministry of Education and Science
of Russian Federation, project No 3.1386.2017, JINR Heisenberg-Landau Program of collaboration Oldenburg-Dubna,
and DFG (Grant LE 838/12-2).

\begin{small}

\end{small}

\end{document}